\documentclass[
superscriptaddress,
showpacs,
nofootinbib,
twocolumn,
aps,
prc,
floatfix
]{revtex4-2}

\usepackage{graphicx}
\usepackage{dcolumn}
\usepackage{bm}
\usepackage{siunitx}
\usepackage[separate-uncertainty=true]{siunitx}
\usepackage{hyperref}
\usepackage{multirow}
\usepackage{ulem}
\usepackage{xcolor}

\hypersetup{
    colorlinks=true,
    linkcolor=blue,
    citecolor=blue,
    filecolor=magenta,      
    urlcolor=cyan,
    }
\usepackage{todonotes}
\usepackage{placeins}
\usepackage{svg}
\DeclareSIUnit\year{yr}
\usepackage{orcidlink}

\newcommand\extrafootertext[1]{%
    \bgroup
    \renewcommand\thefootnote{\fnsymbol{footnote}}%
    \renewcommand\thempfootnote{\fnsymbol{mpfootnote}}%
    \footnotetext[0]{#1}%
    \egroup
}

\begin{document}

\title{Characterization of Low-energy Ionization Signals in Silicon Detectors for the Nab Experiment}

\author{R. J. Taylor}
\affiliation{Department of Physics, North Carolina State University, Raleigh, North Carolina 27695, USA}
\affiliation{Triangle Universities Nuclear Laboratory, Durham, North Carolina 27710, USA}

\author{August Mendelsohn}
\affiliation{Department of Physics and Astronomy, University of Manitoba, Winnipeg, Manitoba R3T 2N2, Canada}

\author{Arlee Shelby}
\affiliation{Department of Physics, North Carolina State University, Raleigh, North Carolina 27695, USA}
\affiliation{Triangle Universities Nuclear Laboratory, Durham, North Carolina 27710, USA}

\author{William C. McCray}
\affiliation{Department of Physics, North Carolina State University, Raleigh, North Carolina 27695, USA}
\affiliation{Triangle Universities Nuclear Laboratory, Durham, North Carolina 27710, USA}

\author{Jin Ha Choi}
\affiliation{Department of Physics, North Carolina State University, Raleigh, North Carolina 27695, USA}
\affiliation{Triangle Universities Nuclear Laboratory, Durham, North Carolina 27710, USA}

\author{Nicholas Macsai}
\affiliation{Department of Physics and Astronomy, University of Manitoba, Winnipeg, Manitoba R3T 2N2, Canada}

\author{Grant Riley}
\affiliation{Los Alamos National Laboratory, Los Alamos, New Mexico, 87545, USA}

\author{Erick Smith}
\affiliation{Los Alamos National Laboratory, Los Alamos, New Mexico 87545, USA}

\author{Stefan Bae{\ss}ler}
\affiliation{Department of Physics, University of Virginia, Charlottesville, Virginia 22904, USA}
\affiliation{Physics Division, Oak Ridge National Laboratory, Oak Ridge, Tennessee 37831, USA}

\author{Leah J. Broussard}
\affiliation{Physics Division, Oak Ridge National Laboratory, Oak Ridge, Tennessee 37831, USA}

\author{Christopher B. Crawford}
\affiliation{University of Kentucky, Lexington, Kentucky 40506, USA}

\author{Michael Gericke}
\affiliation{Department of Physics and Astronomy, University of Manitoba, Winnipeg, Manitoba R3T 2N2, Canada}

\author{Francisco M. Gonzalez}
\affiliation{Physics Division, Oak Ridge National Laboratory, Oak Ridge, Tennessee 37831, USA}

\author{David Harrison}
\affiliation{Department of Physics and Astronomy, University of Manitoba, Winnipeg, Manitoba R3T 2N2, Canada}

\author{Leendert Hayen}
\affiliation{LPC Caen, ENSICAEN, Université de Caen, CNRS/IN2P3, Caen, France}
\affiliation{Department of Physics, North Carolina State University, Raleigh, North Carolina 27695, USA}

\author{Mark Makela}
\affiliation{Los Alamos National Laboratory, Los Alamos, New Mexico, 87545, USA}

\author{Russell Mammei}
\affiliation{Department of Physics and Astronomy, University of Manitoba, Winnipeg, Manitoba R3T 2N2, Canada}
\affiliation{Department of Physics, University of Winnipeg, Winnipeg, Manitoba R3B 2E9, Canada}%

\author{David G. Mathews}
\affiliation{Physics Division, Oak Ridge National Laboratory, Oak Ridge, Tennessee 37831, USA}

\author{Dinko Po{\u c}ani{\'c}}
\affiliation{Department of Physics, University of Virginia, Charlottesville, Virginia 22904, USA}

\author{Glenn Randall}
\affiliation{Department of Physics, Arizona State University, Tempe, Arizona 85287, USA}

\author{Americo Salas-Bacci}
\affiliation{Department of Physics, University of Virginia, Charlottesville, Virginia 22904, USA}

\author{W. Scott Wilburn}
\affiliation{Los Alamos National Laboratory, Los Alamos, New Mexico, 87545, USA}

\author{A. R. Young}
\affiliation{Department of Physics, North Carolina State University, Raleigh, North Carolina 27695, USA}
\affiliation{Triangle Universities Nuclear Laboratory, Durham, North Carolina 27710, USA}

\date{\today}

\begin{abstract}

The Nab (Neutron a b) experiment is designed to measure the beta-antineutrino angular correlation in free neutron $\beta$ decay with an ultimate precision goal of 0.1\%, providing input for tests of Cabibbo-Kobayashi-Maskawa (CKM) matrix unitarity.
This measurement is performed via detection of electrons and protons in delayed coincidence using custom large-area segmented silicon detectors. 
We present the characterization of one such detector system to establish the proton energy and timing response, using a dedicated proton accelerator.

The detected proton peak was studied for \SI{25}{\kilo e\volt}, \SI{30}{\kilo e\volt}, and \SI{35}{\kilo e\volt} incident protons on a set of detector segments and multiple cooling cycles over a one year period.
Ionization losses were consistent with models of the detector dead layer with thicknesses less than \SI{100}{\nano\meter}. 
The detected proton peak was stable within the uncertainty from energy calibration (\SI{0.25}{\kilo e\volt}). 
The rise times of detector pulses from $^{109}$Cd and $^{113}$Sn conversion electron sources were used to extract the impurity density profile and establish a precise model for the detector timing response.
The observed impurity density profile varied from \SI{2\pm2e9}{\per\cm\cubed} at the center to \SI{26\pm2e9}{\per\cm\cubed} at the edge.
This impurity density profile was then used to characterize systematic effects in proton time-of-flight measurements due to detector pulse-shape effects; the resultant proton timing systematic uncertainties were below \SI{0.3}{\nano\second}, which is sufficient for the Nab experiment. 

\end{abstract}


\maketitle

\extrafootertext{Notice: This manuscript has been authored in part by UT-Battelle, LLC, under contract DE-AC05-00OR22725 with the US Department of Energy (DOE). The US government retains and the publisher, by accepting the article for publication, acknowledges that the US government retains a nonexclusive, paid-up, irrevocable, worldwide license to publish or reproduce the published form of this manuscript, or allow others to do so, for US government purposes. DOE will provide public access to these results of federally sponsored research in accordance with the DOE Public Access Plan (https://www.energy.gov/doe-public-access-plan).}

\section{Introduction} \label{sec:intro}

Precise measurements of $\beta$ decay provide rigorous tests that refine our understanding of the Standard Model of Particle Physics. One such test involves measurement of elements of the  Cabibbo-Kobayashi-Maskawa (CKM) matrix, which is unitary in the Standard Model. 
There is currently tension in measurements of CKM unitarity as determined from kaon and superallowed $0^+ \rightarrow  0^+$ decays, referred to as the Cabibbo Angle Anomaly (CAA) \cite{ciriglianoScrutinizingCKMUnitarity2023, Antonelli:2009ws}. 
Precision measurements of free neutron decay can provide valuable input in determining $V_{ud}$, since they do not include nuclear correction uncertainties present in superallowed decays, therefore targeting one possible contribution to the CAA. 
However, determining $V_{ud}$ from neutrons does require a separate measurement of the nucleon isovector axial charge, $g_A$, which can be determined from a measurement of the electron-neutrino angular correlation.

The decay rate, $\Gamma$, for unpolarized neutrons is given to leading order by \cite{Jackson1957}
\begin{equation}\label{eq:decay_rate}
    \Gamma \propto 1 + a \frac{\vec{p}_e \cdot \vec{p}_{\bar{\nu}_e}}{E_e E_{\bar{\nu}_e}} + b\frac{m_e}{E_e},
\end{equation}
where $\vec{p}_e$ and $E_e$ are the momentum and energy of the electron, $\vec{p}_{\bar{\nu}_e}$ and $E_{\bar{\nu}_e}$ are the momentum and energy of the neutrino, the parameter $a$ specifies the electron-anti-neutrino angular correlation, and $b$ is the Fierz interference coefficient (zero in the Standard Model of particle physics).
Because direct detection of the anti-neutrino is not feasible, to determine $a$ the Nab experiment relies on momentum and energy conservation to write the observables in terms of the proton momentum $p_p$ and the electron energy, to leading order
\begin{equation}\label{eq:kinematics}
    \Gamma \propto 1+a\frac{p_p^2-p_e^2-(E_0-E_e)^2}{2E_e(E_0-E_e)} ,
\end{equation}
where $E_0$ is the electron endpoint energy.
The Nab experiment is designed to measure $a$ with an ultimate precision goal of 0.1\% \cite{pocanic_2009, Nab:2012hnc}.
A second physics target for Nab involves  measurements of the electron energy spectrum to place sub-percent constraints on the presence of the Fierz interference term (Eq.~\ref{eq:decay_rate}) \cite{Baessler_2024}.

In the Nab experiment, detection of the emitted electrons and protons following neutron decay is accomplished in silicon detectors positioned at either end of an asymmetric magnetic spectrometer.  
The emitted proton's momentum is determined through a time-of-flight measurement in one arm of the spectrometer, where the momentum is first adiabatically longitudinalized before the proton enters a five meter long drift region to provide high sensitivity to the time-of-flight.  
Electron events are hermetically contained between the detectors at either end of the spectrometer and occur over a much shorter time interval following the decay, establishing the ``start" of a coincidence timing event with emitted protons.  
Due to the low proton energy  following beta decay (less than \SI{751}{e\volt}), protons in the Nab experiment are accelerated to approximately \SI{30}{\kilo e\volt} prior to striking the detector at the end of the drift region. This acceleration ensures a very high probability for protons to penetrate the Si detector entrance window and deposit energy above the detection threshold.

Beta decay experiments have integrated Si detectors into the experimental approach to refine particle energy and event timing extraction \cite{pocanic_2009,urbanCharacterizationMeasurementsTRISTAN2022,mertensNovelDetectorSystem2019,BL3_2014}.
The Nab experiment utilizes 1.5 to \SI{2}{\milli\meter} thick, highly segmented Si detectors approximately \SI{12}{\centi\meter} in diameter.
The excellent linearity and precise timing expected for Si detectors were strong drivers behind their adoption for the Nab experiment.
These detectors are fabricated with entrance window dead layers up to \SI{100}{\nano\meter} thick. 
The thin dead layer is required to ensure essentially all protons are above the energy threshold for detection.   

To optimize the Nab detector performance, an assessment of the expected response to protons was performed using a low-energy accelerator.
Previous investigations of Nab detectors focused on the energy response of a similar, single segment of a detector to protons \cite{Salas-Bacci2014} and a first investigation of the response using the proton acceleration scheme planned for Nab \cite{Broussard2017_2, Broussard2017}. 

In this work we describe: the driving constraints for detector performance (Section~\ref{sec:nab_requirements}), an overview of the characterization apparatus and measurements made (Section~\ref{sec:char_overview}), energy calibration of the silicon detector (Section~\ref{sec:detector_response}), the measured proton energy spectra and a comparison with a simulated response (Section~\ref{sec:proton_charac}), an analysis of pulse shapes to determine the expected performance of these detectors for proton timing measurements (Section~\ref{sec: Waveform Analysis}), and finally the impact of our characterization effort for the Nab experiment (Section~\ref{sec:operating_conditions}).

\section{Nab Detector Requirements}
\label{sec:nab_requirements}
The two quantities of interest in this work are, for a given decay event, the integrated charge produced by incident protons on a given segment (or pixel) and the time evolution of the signal pulse.
Charge-sharing is possible near the boundaries between two pixels and the time evolution of ionization signals varies strongly with the location of the incident beam on a given pixel \cite{Hayen2023}.
To provide direct and reliable comparisons with expected proton ionization signals on a representative set of pixels, we used the University of Manitoba proton source facility to provide steerable, mono-energetic beams of \SI{30}{\kilo e\volt} protons \cite{macsai2025}.
Since pixels have an effective diameter of about \SI{1}{\centi\meter}, these requirements were met with characteristic beam diameters of roughly \SI{2}{\milli\meter} and positioning precision also of about \SI{2}{\milli\meter}.
This capability facilitated one of the primary goals of this work: the reliable detection of expected proton signals near the center of a selected pixel.
To satisfy Nab operational requirements, characterization of the proton energy signals included a comparison with the expected energy deposition for several possible dead-layer models and over repeated cryogenic thermal cycling.

For Nab, reliable timing for both proton and electron pulses is required to extract the angular correlation parameter \cite{gonzalezFirstFullDalitz2025}. 
The preferred Nab analysis, method ``B" described in the Nab proposal, demands that any systematic error in the relative time interval between an electron ``start" and a proton ``stop" be less than about \SI{0.3}{\nano\second} \cite{Fry_2019,alarconPreciseMeasurementNeutron,Hayen2023}. 
We define any such uncorrected systematic error as a time of flight bias.
Therefore to achieve the ultimate sensitivity for Nab we must determine the correction for any possible timing bias with an uncertainty less than \SI{0.3}{\nano\second}.

In order to ascertain the correction for the timing bias, the timing properties of proton and electron pulses must be defined. 
Timing is typically accomplished through the determination of a particular point on the pulse produced by the particle, such as the point at which the pulse is greater than a fixed voltage threshold or when the pulse reaches some fixed fraction (such as 50\%) of its maximum amplitude.  
Each particular timing algorithm has a characteristic ``timing offset" between the true start of the pulse and the time identified by the algorithm.
Because the detector response is different for electrons and protons, knowledge of a potential offset for any selected timing algorithm is essential for both electron and proton events. 

Measuring the timing biases is complicated by an expected radial variation of the silicon impurity density profile which introduces radially dependent charge transport times. 
Charge transport times affect signal shape, so this radial variation will result in a pixel specific timing response \cite{Hayen2023}.
A set of mono-energetic electron sources are required in addition to the proton source in order to determine the detector impurity density profile and evaluate potential timing offsets.
The radioactive sources used for detector energy calibration, $^{109}$Cd and $^{113}$Sn, proved suitable for this analysis.

In this work, we document the procedures used to characterize the energy signal and the expected variation in proton pulse detection under operating conditions nearly identical to those for the Nab experiment.
Further, we demonstrate that accurate characterization of the expected average timing offsets can be predicted through empirically constrained pulse models.  

\section{Apparatus and Measurement Overview}
\label{sec:char_overview}

\begin{figure}
\includegraphics[width=0.48\textwidth]{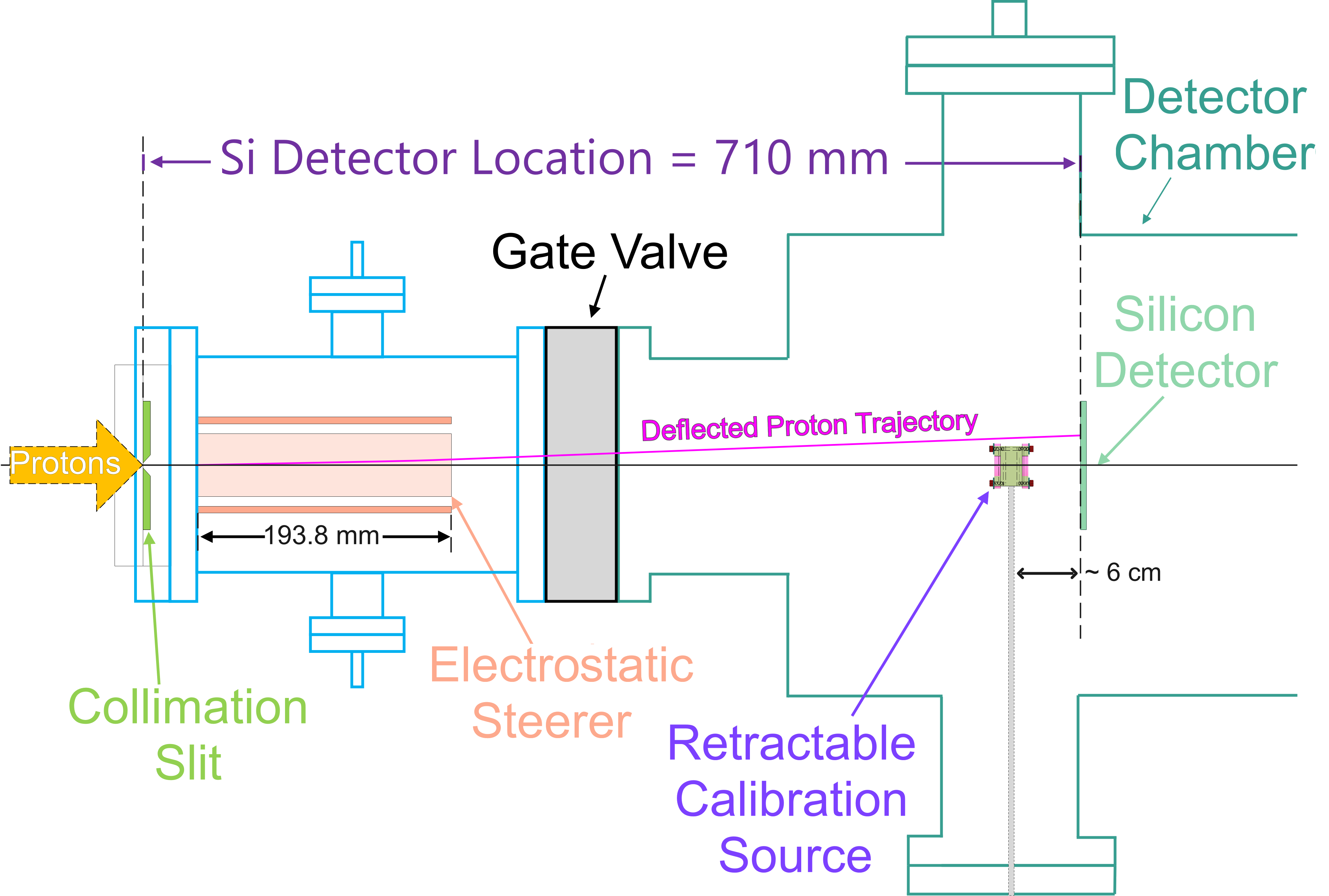}
    \caption{Top-down view of the vacuum vessel containing the detector. A retractable source arm can be rotated to expose one of two radioactive sources, or the collimated beam of protons can directed to the desired pixel using the electro-static steerer.}
    \label{fig:detector_region}
\end{figure}

Characterization of the detector was performed with a beam of \SI{30}{\kilo e\volt} protons and $^{109}$Cd and $^{113}$Sn radioactive sources that could be positioned near the detector entrance window (see Fig.~\ref{fig:detector_region}).
Here we describe the working principles of the detector system, the proton source used for these measurements, and calibration sources used to reach Nab's precision goals in energy resolution and timing bias.

\subsection{The Nab Detection System}
\label{subsec:Nab_silicon}

The Nab silicon detectors are described extensively in Ref.~\cite{Hayen2023}, however we briefly summarize relevant features of the detection system here.

\subsubsection{Nab Silicon Detectors}
\label{subsubsec:silicon_diode_detectors}

\begin{figure}
\includegraphics[width=0.48\textwidth]{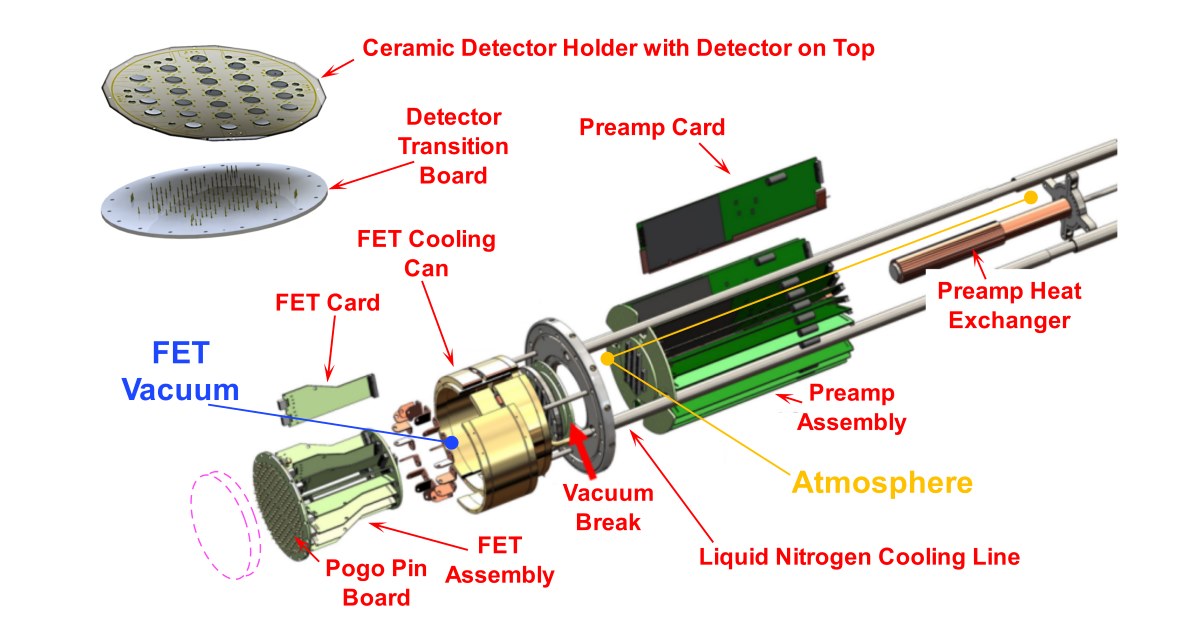}
    \caption{The silicon detector is affixed to the front of an assembly referred to as the ``detector system,'' the internals of which are depicted above. The detector system is composed of the JFET and amplifier assemblies, the amplifier liquid cooling, and the cryogenic detector cooling system. Each amplifier card contains six channels (amplifier circuits for six detector pixels - see Fig.~\ref{fig:spice_model}). There are 22 power cables in addition to the 127 coaxial signal cables (not shown). The dashed lines at the bottom left indicate the mounting position for the detector and detector transition board. This assembly is installed in a stainless steel tube to separate the FET vacuum from the ultra-high vacuum detector volume.}
    \label{fig:internal_electronics}
\end{figure}

The silicon detectors for the Nab experiment consist of \SI{1.5}{\milli\meter} or \SI{2}{\milli\meter} thick, \SI{150}{\milli\meter} diameter wafers, with an active area of \SI{368}{\milli\meter\squared} segmented into 127 pixels.
The $n$-type bulk is implanted with boron $p^+$ ions to create the diode junction, and the entrance window.
An aluminum grid is sputtered on to the entrance window to enable uniform reverse-biasing of the $p$-$n$ junction while only covering 0.4\% of the window \cite{Broussard2017}.
The reverse, or ohmic side, is implanted with phosphorus $n^+$ ions in hexagonal pixels to create the segmentation.
Once the segments are created, the surface is metallized to create a conducting contact with the detector electronics system.

The detectors are fabricated by Micron Semiconductor Ltd (\cite{Micron}), employing a p-spray method that involves uniformly implanting boron over the surface, then localized implantation of phosphorus to create the $n^+$ pixels.
The detector used for the proton studies at the University of Manitoba described here was a \SI{2}{\milli\meter} thick wafer.

\subsubsection{Amplifier and Electronics}
\label{subsubsec:Amplifier}

\begin{figure*}
    \centering
    \includegraphics[width=\textwidth]{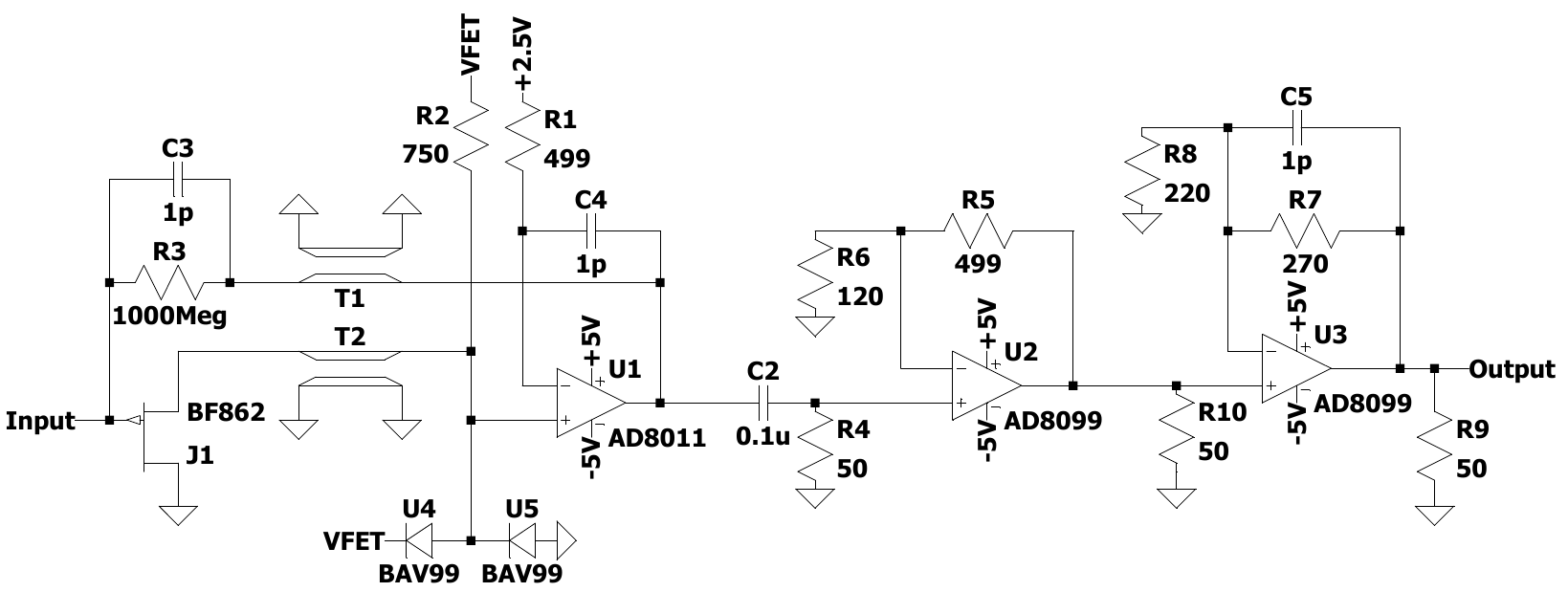}
    \caption{The front-end amplifier circuit modeled in LT-SPICE \cite{ltspice}. The trans-impedance amplifier is composed of a common-source BF862 JFET (J1) coupled to an AD8011 video amplifier (U1). The detector, JFET, and feedback components (C3, R3) are all cooled via LN$_2$. T1 and T2 represent the ``transition boards'' that feed signals between the warm and cold sections. Following the trans-impedance stage, there is an integrator (C2, R4) followed by two amplification stages that act as differentiators (U2, U3). U4 and U5 are protection diodes for the input bias to U1. R9 and R10 are in place for impedance matching.}
    \label{fig:spice_model}
\end{figure*}

Each pixel contact on the detector is coupled to a cryogenically-cooled junction field-effect transistor (JFET) in-tandem with a current-feedback amplifier. 
This configuration acts as a trans-impedance pre-amplifier which precedes two further gain stages.
Figure \ref{fig:spice_model} shows the schematic for one pixel's amplification chain.
The BF862 JFET (J1) was chosen to match the pixel capacitance, and the AD8011 pre-amplifer (U1) was selected based on its fast performance.
The amplifiers are placed in groups of six on printed circuit boards (PCBs), and the JFET elements in groups of eight; this configuration means that pixels also receive power in groups of six (see Fig.~\ref{fig:internal_electronics}).
The pulses are then sent by coaxial cable to National Instruments PXIe-5171R FPGA analog-to-digital converters (ADCs). This data acquisition system digitizes pulse waveforms and saves the corresponding meta-data to compressed hierarchical data format files (HDF5) \cite{MATHEWS2025170079,Mathews2024}.

High gain signals fed back to the JFET coupled with fast pulse edges meant that instability was unavoidable.
An additional effect suspected to have contributed to instability was the complex impedances inside the interconnect PCBs between the warm and cold sections of the amplifiers \cite{Broussard2017_2}.
This had the potential to cause oscillatory output in a neighboring set of pre-amplifiers, cascading through the circuits instrumented for the remaining pixels.
As a result, we instrumented a subset of the pixels that kept the system operable, while still enabling us to study effects radially from the center (see Fig.~\ref{fig:powered_pixels}).
For all of the analysis included in this publication, the instrument configuration was kept constant except for the detector bias voltage ($V_b$) and the temperature of the components from the detector to the vacuum break (see Fig.~\ref{fig:internal_electronics}).

\subsubsection{Detector Cooling}
\label{subsubsec:detectormount}

The detector system contains the detector, and front-end amplifiers for each segment, as well as the necessary provisions for cooling (Fig.~\ref{fig:internal_electronics}).
The detector and JFET were cooled using liquid nitrogen (LN$_2$), regulated with a mass-flow controller.
Cryogenic cooling drastically reduces the diode-junction leakage current, and also improves the JFET timing performance.
Further, cooling improves the charge collection time as a power-law with temperature \cite{Hayen2023}.
However, cooling the JFETs below approximately \SI{110}{\kelvin} impedes performance, so the assembly is optimized to run at or above this temperature \cite{Knoll_2024}.

\subsection{The Manitoba II Proton Source}
\label{subsec:proton_source}

In Nab, the full detector and data acquisition system is electrically referenced to \SI{-30}{\kilo \volt}, which allows the protons to be accelerated to a more detectable energy \cite{gluckMeasurableDistributionsUnpolarized1993}.  
To characterize proton detection at these energies, the Manitoba II mass spectrometer was adapted using a Penning ion generator to form a beam of \SI{30}{\kilo e\volt}, leaving the detector system held at ground potential \cite{Barber1971,Harrison2013,macsai2025}.

The proton source facility includes the ability to steer the proton beam using a set of electro-static plates set up in a rectangular configuration (Fig.~\ref{fig:detector_region}).
Momentum resolution of the magneto-static analyzer was measured to be 1\% \cite{Harrison2013}, corresponding to an energy resolution for the proton beam energy of \SI{300}{e\volt} full-width at half-maximum (FWHM).
The proton beam rate  was approximately \SI{10}{counts\per\second} and operated the detector in a vacuum of $\leq$\SI{5e-7}{Torr}.
The boundary between two pixels was used to determine the proton beam spot size of $\sim$\SI{1\pm 0.1}{\milli\meter}, which meets the requirement of a beam width less than \SI{2}{\milli \meter} for detector characterization \cite{macsai2025}.

\subsection{Radioactive Calibration Sources}
\label{sec:rad_sources}

We measured conversion electron and x-ray peak energies from two sources, $^{113}$Sn and $^{109}$Cd, in order to determine a single pixel-by-pixel calibration model across a wide range of energies. 
We calibrated using two x-ray lines, computed as a weighted average for all K$\alpha$ and K$\beta$ lines, and three conversion electron lines. 
The peaks used for calibrations are the 24.14 keV (K$\alpha$) x-ray, 27.35 keV (K$\beta$) x-ray, \SI{363.76}{\kilo e\volt} electron, \SI{387.46}{\kilo e\volt} electron, and \SI{390.87}{\kilo e\volt} electron peaks from $^{113}$Sn, and the 22.11 keV (K$\alpha$) x-ray, 25.01 keV (K$\beta$) x-ray, \SI{65.52}{\kilo e\volt} electron, \SI{84.23}{\kilo e\volt} electron, and \SI{87.40}{\kilo e\volt} electron peaks from $^{109}$Cd \cite{nndc_sn, nndc_cd}. 
The $^{109}$Cd \SI{87.40}{\kilo e\volt} peak is a weighted average of the 87.3 keV and 87.9 keV electron lines. 

The radioactive sources were prepared by Eckert \& Ziegler Isotope Products with the radioactive isotope in a carrier solution encapsulated by two pieces of aluminized Mylar foil glued together \cite{EckertZieglerReference2007}. 
These sources were mounted on a retractable source arm (Fig.~\ref{fig:detector_region}) and positioned roughly 6 cm in front of the detector for calibration measurements, but removed for the proton beam measurements. 

\begin{figure}
\includegraphics[width=0.4\textwidth]{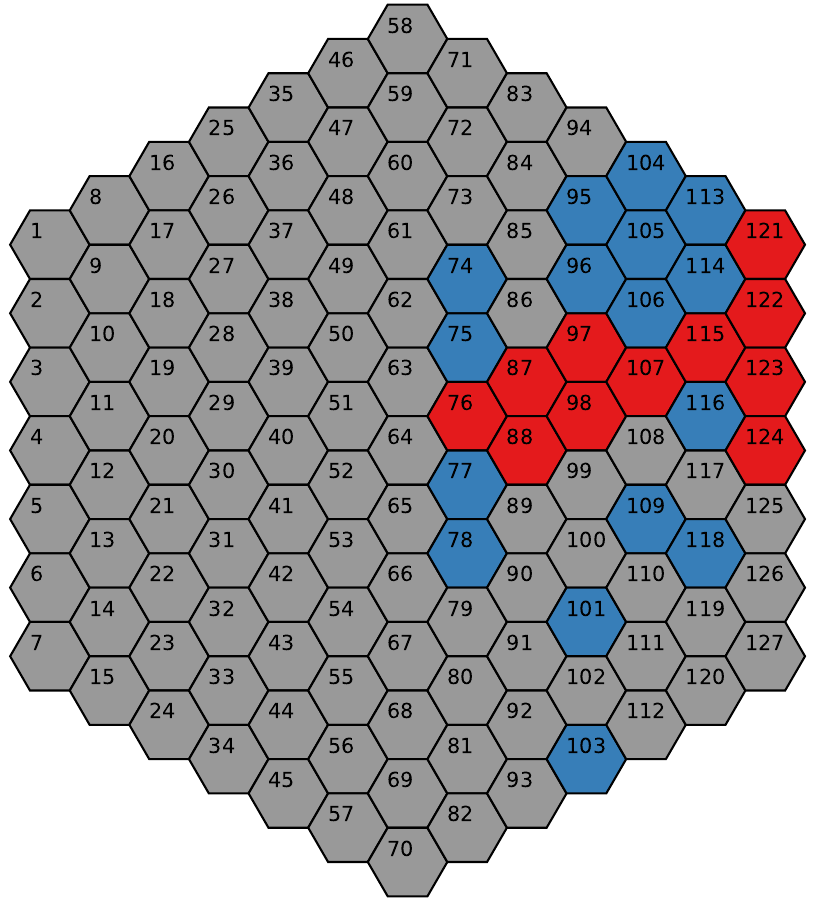}
    \caption{Pixels studied with $^{109}$Cd, $^{113}$Sn, and protons are highlighted in red, whereas pixels only studied with radioactive sources are in blue. Due to amplifier stability constraints, further pixels were not instrumented. The pattern of powered pixels is due to pre-amplifier PCBs having up to six channels per card, and electronic stability constraints (see Sec. \ref{subsubsec:Amplifier}).}
    \label{fig:powered_pixels}
\end{figure}

\subsection{Data Summary}
\label{sec:data_summary}

The analyses presented in this work are based on the studies listed below: 
\begin{itemize}
    \item $^{113}$Sn with detector bias voltage ($V_b$) from \SI{- 300}{\volt} to \SI{- 60}{\volt}, in nine steps.
    \item $^{109}$Cd with $V_b$ from \SI{- 320}{\volt} to \SI{- 10}{\volt}, in 32 steps.    
    \item \SI{30}{\kilo e\volt} protons with $V_b$ from \SI{- 320}{\volt} to \SI{- 100}{\volt}, in 12 steps.    
    \item \SI{25}{\kilo e\volt}, \SI{30}{\kilo e\volt}, and \SI{35}{\kilo e\volt} protons at a $V_b$ of \SI{-300}{\volt}, a range of temperatures (\SI{115}{\kelvin}, \SI{125}{\kelvin}, \SI{135}{\kelvin}, and \SI{145}{\kelvin}), and on a radially distributed set of pixels (76, 98, 121-124 in Fig.~\ref{fig:powered_pixels}).
    \item $^{109}$Cd high-statistics datasets at a range of temperatures between 117 and \SI{150}{\kelvin}.
\end{itemize}

These studies were broken down into individual datasets collected while the detector was at the stated temperatures (\SI{120}{\kelvin} otherwise).
The cooling cycle consisted of using a \SI{220}{\liter} Dewar of LN$_2$ regulated with a mass flow controller. This setup was stable to \SI{2}{\kelvin} and lasted for a period of approximately \SI{4}{days}.

High-statistics calibration with a $^{109}$Cd source was performed  as the first and last activity in a cooling cycle.
$^{109}$Cd was chosen for these routine calibration studies since its decay has an X-ray peak near the detected proton peak and conversion electron peaks at slightly higher energies.
The $^{113}$Sn source provided high signal-to-noise waveforms for calibration and signal rise-time studies.
During data collection we recorded the temperature, voltage, and current draw of the detector and supporting electronics.

\section{Detector Energy Calibration}
\label{sec:detector_response}

To characterize the detector energy response, an optimized trapezoidal filter was used to extract pulse energies \cite{jordanov1994digital}. The shape of the trapezoidal filter was optimized to minimize the noise contributions.
Resultant energy spectra were then compared to ionization signals predicted from Geant4 to determine the calibration for pixels under study.

\subsection{Trapezoidal Filter Optimization}
\label{subsec:Trap_Filter_Study}

A trapezoidal filter is convolved with detector signal pulses to accurately determine the integrated charge while minimizing the impact of noise \cite{Knoll_2024, jordanov1994digital}.
When applied to a pulse, the output of the filter linearly rises to a maximum value over a time interval referred to as the ``shaping time" ($t_S$), remains roughly constant over a period called the ``flat top", and then falls linearly back to zero over a final time interval of length $t_S$.  
The flat top height is proportional to the integrated charge. 
A final input parameter to the filter is the ``decay time" of the signal output from the preamplifier.
A study was performed to determine the optimal trapezoidal filter parameters minimize the effect of noise in this detector.

For each pixel, the trapezoidal filter parameters were varied and a fit applied to the $^{113}$Sn \SI{364}{\kilo e\volt} electron peak. 
An optimum flat-top time of \SI{0.3}{\micro\second} was determined, providing adequate time for charge collection.
The optimum pulse fall time correction was determined by varying the fall time parameter until a minimal slope for the signal amplitude (detected charge) was achieved.
The fall time correction agrees with the nominal \SI{5}{\micro\second} at the percent level due to component manufacturing variations. 

\begin{figure}
    \centering
    \includegraphics{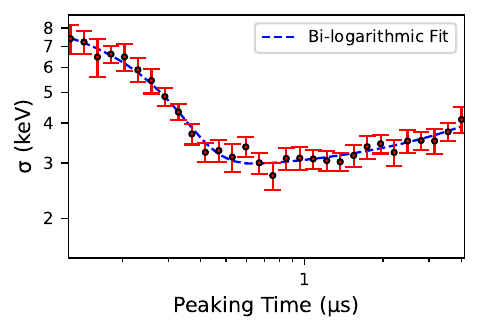}
    \caption{An example of the energy standard deviation ($\sigma$) versus peaking time for the $^{113}$Sn \SI{364}{\kilo e\volt} conversion electron peak. The data is fit to a bi-logarithmic function: $y=\ln(Ae^{-x/B}+Ce^{x/D})$. The minimum of the bi-logarithmic fit (shown in blue) indicates the minimal noise contribution to the trapezoidal filter energy extraction. Therefore it is used to set the peaking time of the filter algorithm.}
    \label{fig:sigma_peak_time_2}
\end{figure}

The measured Gaussian width parameter, $\sigma$, for mono-energetic signals is proportional to the equivalent noise charge (ENC) and results from series and parallel noise \cite{Bertuccio1993}.
An optimum peaking time was found for each pixel that minimizes $\sigma$ and therefore both sources of noise.
For pixel 76, the best peaking time was \SI{0.82}{\micro\second}, corresponding to a full-width at half-maximum (FWHM) of \SI{6.9}{\kilo e\volt}  (Fig.~\ref{fig:sigma_peak_time_2}).

The mean optimal peaking-time for all pixels analyzed was \SI{0.56}{\micro\second} with a standard deviation of \SI{0.20}{\micro\second}. 
The mean optimum $\sigma$ in the analysis was \SI{2.92}{\kilo e\volt} with a standard deviation of \SI{0.41}{\kilo e\volt}, corresponding to a FWHM of \SI{6.9 \pm 0.97}{\kilo e\volt}. 
The noise was dominated by radiative pick-up and ground-loops -- leakage current did not make significant contributions to the parallel noise.  
Although further reduction in the electronic noise has been implemented by the Nab collaboration since these data were obtained, the observed noise performance was sufficient for the goals of this work.
A more detailed analysis of the noise parameters for an earlier version of the electronics chain can be found in Ref. \cite{Broussard2017_2}.

\subsection{Integral Cross Talk Assessment}
\label{subsec:Integral_Crosstalk}

When energy is deposited in a given pixel, small signals are typically induced in the neighboring pixels, referred to as ``cross-talk".
Cross-talk between pixels in the front-end of the detection system has potential to impact the optimization of the proton trigger threshold in Nab. 
As explained in Section~\ref{subsubsec:detectormount}, the decay protons are accelerated to \SI{30}{\kilo e\volt} as a way to improve detection. 
However, cross-talk imposes a lower limit on the pulse trigger threshold, and the resolvability of the proton energy peak from noise and accidental triggers.

There are two contributions to cross-talk. 
Differential cross-talk arises from the movement of charge carriers in adjacent pixels during the charge collection time and does not result in an integrated charge being measured.
The integral contribution arises mainly from the mutual capacitance of the readout electrode geometry, with some contribution from the pre-amplifier electronics \cite{Cornat_2009,Pullia_2011}.

To study integral cross-talk, the large-amplitude signals from the $^{113}$Sn \SI{364}{\kilo e\volt} conversion electron peak were used as a primary trigger to look for induced signals on neighboring pixels.
Since the effect is linear in energy, the results from the $^{113}$Sn study can be extended to predict cross-talk for proton signals expected at lower energies \cite{Knoll_2024}. 
The DAQ was configured to digitize waveforms from each pixel and the six adjacent pixels simultaneously. 
Each event consists of seven waveforms with the same event tag, differentiated by a primary or secondary tag. 
A few percent of the waveforms with a transient differential cross-talk feature were cut from the average to isolate effects due to the integral cross-talk.
These waveforms contained transient components during charge collection much larger than the integral cross talk amplitude (on the order of 30\% that of the original electron pulse), and therefore interfered with pulse averaging.

\begin{figure}
    \includegraphics{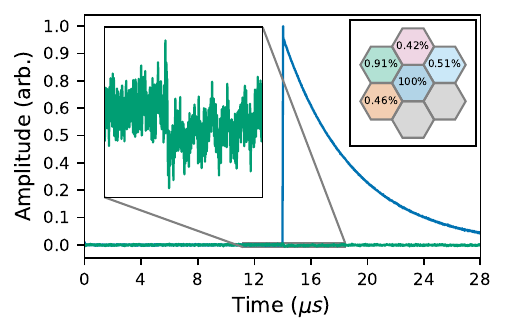}
    \caption{Example of correlated cross-talk between neighbor detector pixels. $^{113}$Sn \SI{364}{\kilo e\volt} conversion electron  events (blue) were compared to the induced cross-talk (one example is in green).The inset at top-right shows the comparison of extracted energies when convolved with a trapezoidal filter. Note that the greyed out pixels were not studied.}
    \label{fig:cross_talk}
\end{figure}

For a given pair of neighbor pixels, waveforms with a secondary tag were averaged (to improve signal-to-noise ratio) and the trapezoidal filter was used to extract the energy, and compared to the averaged primary-tagged waveforms on the central pixel. 
An example of this is shown in Fig.~\ref{fig:cross_talk}; the percentages presented in the figure are the ratio of the extracted energy after convolution with a trapezoidal filter, referred to as the cross-talk energy ratio.

A group of twenty detector pixels were included in the study, resulting in a cross-talk energy ratio less that approximately $1\%$. 
This meets the operating requirements for the Nab experiment and is consistent with studies found in the literature \cite{MATHEWS2025170079,Cornat_2009,Pullia_2011}.

\subsection{Calibration with Radioactive Sources}
\label{subsec:Source_Calibration}

\begin{figure*}
    \includegraphics{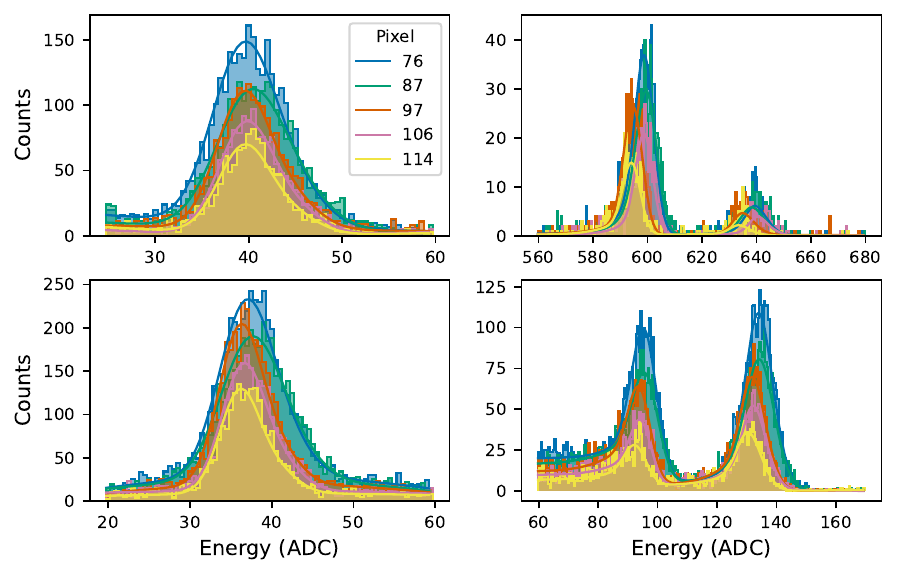}
    \caption{The fit results for $^{113}$Sn and $^{109}$Cd source data, with no cuts applied to the measured data set, for five radial pixels as indicated in the legend. The top two plots correspond to the $^{113}$Sn x-ray (left) and conversion electron lines (right). The bottom two plots correspond to the $^{109}$Cd x-ray (left) and conversion electron lines (right).}
    \label{fig:calibration_fits}
\end{figure*}

One of the primary objectives for the data collected at Manitoba was to make an assessment of the proton response of the Nab detectors.
To establish the energy calibration for the proton signals, Geant4 simulations were developed to model the measured energy spectra from two radioactive sources: $^{113}$Sn and $^{109}$Cd \cite{AGOSTINELLI2003250,ALLISON1610988}.  
These were used to calibrate the spectra measured using the optimized trapezoidal filter in Section~\ref{subsec:Trap_Filter_Study}.
Our calibration procedure resulted in overall uncertainties of roughly 0.2 keV for the ionization signals for incident proton beams, providing a sensitive tests, for example, for the expectations from various dead-layer models.

For conversion electron spectra, we used the class of fitting functions from Ref.~\cite{LONGORIA1990308}.
In general, the electron energy lines were fit to a linear background and two peaks, each consisting of a Gaussian, a lower-lying exponential, and a step function. The higher energy peak contains two unresolvable energy lines, so another Gaussian was included, constrained by relative energy and intensity ratios.
For x-ray spectra, we fit the sum of two Gaussians that are constrained by energy and intensity ratios of K$\alpha$ and K$\beta$ peaks and a second order polynomial background model. 
We show the results in Fig.~\ref{fig:calibration_fits} for five pixels at $V_b$ = \SI{-300}{\volt}, that extend radially out from the center (76, 87, 97, 106, and 114 as labeled in Fig.~\ref{fig:powered_pixels}).

From the obtained conversion electron and x-ray peak centroids (in arbitrary ADC units), we calibrated individual detector pixels with six energy lines, three from $^{113}$Sn and three from $^{109}$Cd (Sec.~\ref{sec:rad_sources}). 
In order to include the fit results from both sources in a single calibration model, we corrected for temperature differences between the corresponding $^{113}$Sn and $^{109}$Cd datasets and corrected the conversion electron lines due to energy loss through the source foils using a Geant4 simulation. 
The corrections and results are described in the following subsections.

\subsubsection{Temperature Correction}
\label{subsubsec:temp_correct}

In order to analyze the temperature dependence of the Nab detectors, $^{109}$Cd data were taken at $V_b=\SI{-300}{\volt}$ for three different temperatures: \SI{124}{\kelvin}, \SI{133}{\kelvin}, and \SI{151}{\kelvin}. 
In our temperature model, we assumed a linear decrease in ionization energy, $\epsilon_{ph}$, as a function of temperature for Si \cite{PEHL196845}.  
A small dependence on the temperature is also expected from the detector electronics. 
From $\epsilon_{ph} = \frac{E}{\langle N \rangle}$, where E is the incident radiation energy and N is the number of electron-hole pairs created, we estimate a temperature correction from the ratio, 

\begin{equation}\label{eq:eph_ratio}
    \frac{\epsilon_{ph,2}}{\epsilon_{ph,1}} = \frac{\langle N_{1} \rangle}{\langle N_{2} \rangle} \propto \frac{E_{1}}{E_{2}},
\end{equation}


at a particular incident energy. Here, $E_{1}$ and $E_{2}$ are estimated by the peak centers obtained at $T_{1}$ and $T_{2}$. 
To analyze the validity of this temperature model, we extracted $\epsilon_{ph}$ at the above temperatures from \cite{PEHL196845} and compare ratios of $\epsilon_{ph}$ to ratios of the extracted peak centers. 
We also constrained the slope of the temperature dependence using uncertainties from \cite{PEHL196845}. 
We found that this temperature model was consistent with the data within the uncertainties in the fit peak centers. 
Additionally, a small dependence on temperature is expected in the detector electronics response, but the uncertainties in the data were too large to observe this.
Our statistical uncertainties were too large to stringently test whether the assumed linear energy dependence of the ionization energy was sufficient to capture the temperature dependence of the calibration.
This temperature dependent energy correction was included in all subsequent analyses.

\subsubsection{Radioactive Source Event Simulation}
\label{sec:source-event-simulation}

The radioactive source configuration consists of two layers of aluminized Mylar foil that are glued together to encapsulate the radioactive material suspended in a carrier compound.
Emitted source electrons have energy dependent losses through the foil that must be accounted for in the energy calibration.
To do so, the source was modeled in Geant4 using the approximate specifications provided by the source manufacturer \cite{EckertZieglerReference2007}.

The Geant4 simulations include the physical geometry of the silicon detector and detailed representations of the radioactive sources.
These simulations provide an expectation for the energy spectroscopy of the calibration sources and the ionization profile within the silicon detector \cite{AGOSTINELLI2003250,ALLISON2016186,ALLISON1610988}. 
The emission of decay products of the radioactive sources are simulated in full cascade: both $^{113}$Sn and $^{109}$Cd decay via the electron capture process, and they emit a chain of conversion and Auger electrons in cascade along with x-rays and gamma rays, dictated by the decay branch and conversion efficiencies. 
Geant4 then transports the particles through the radioactive source structures and to the silicon detector. 

\begin{figure}
    \includegraphics{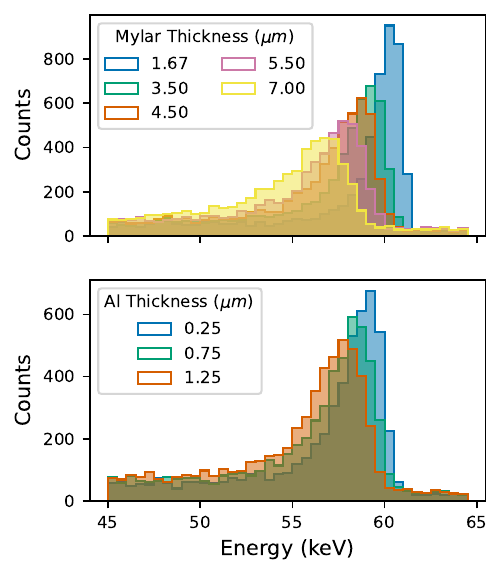}
    \caption{\SI{63}{\kilo e\volt} $^{109}$Cd peak energy histograms from the Geant4 simulations for pixel 76. The top plot shows the results as Mylar thickness is changed, with Al and effective carrier material thicknesses kept constant at \SI{1}{\micro\meter} and \SI{20}{\nano\meter} respectively. The bottom shows the histograms as Al thickness is changed, with the Mylar and carrier material thicknesses kept constant at \SI{5}{\micro\meter} and \SI{20}{\nano\meter} respectively.}
    \label{fig:GEANT_mylar_Al_histograms}
\end{figure}

An example of the simulated electron energy response is shown in Fig.~\ref{fig:GEANT_mylar_Al_histograms} for pixel 76 for the \SI{63}{\kilo e\volt} $^{109}$Cd peak. The top plot shows the peak shifting to lower energies and broadening as a function of increasing the Mylar thickness and the bottom plot shows how the peak shifts to lower energies as a function of increasing the aluminum thickness.

\subsubsection{Energy Loss Correction from Event Simulation}
\label{sec:energy-loss-correction}

We expect energy losses for conversion electrons due to the source configuration and must apply a correction for both $^{113}$Sn and $^{109}$Cd.
We expect losses to be significant for the $^{109}$Cd conversion lines, as they are much lower in energy compared to the $^{113}$Sn lines. 
We also expect losses to be radially dependent on the pixel location on the detector.
Thus, we determined the most likely source configuration from five radial pixels: 76, 87, 97, 106, and 114 as shown in Fig.~\ref{fig:powered_pixels}.
A set of Geant4 simulations, as described previously, were performed for source configurations with a Mylar thickness ranging from 1.5 - \SI{27.0}{\micro\meter} in \SI{0.5}{\micro\meter} steps and an aluminum thickness ranging from 0.25 - \SI{5.0}{\micro\meter} in \SI{0.25}{\micro\meter} steps. 
The set also included 6 different effective carrier material thicknesses; 1, 5, 15, 20, 40, and \SI{100}{\nano\meter}.

To determine the most likely source configuration, we used a 68\% confidence interval estimation method, which included two kinds of fits for every simulated source configuration: a ``spectrum fit" and a ``peak fit". 
The spectrum fit has 7 fit parameters to minimize the difference between the simulated and measured spectra. 
The fit procedure includes convolving the simulated spectrum with a Gaussian, shifting the simulated energy axis to the measured axis, an interpolation onto the shifted axis, normalization, and quadratic background subtraction. 
The peak fit uses the fitting functions from Sec.~\ref{subsec:Source_Calibration} to obtain corrected energy values in \SI{}{\kilo e\volt} for the energy lines used in the calibration scheme, from the convolved simulated spectrum.

For each source, we used the spectrum fit results to obtain a $\chi^{2}$ distribution as a function of Mylar and aluminum thickness, using a summed value representing the fit results for the five pixels together. 
For $^{109}$Cd, the most likely effective carrier material thickness was determined by the distribution corresponding to the smallest minimum $\chi^2$.
Due to the limiting statistics of the $^{113}$Sn data, we were not sensitive to the effective carrier material thickness and assumed a most likely value consistent with the manufacturer specifications and the $^{109}$Cd results.

Fixing the effective carrier thickness to the selected values for each source, we use the associated $\chi^{2}$ distribution to determine the 68\% confidence interval region, and fit it to an ellipse constrained to be centered at the minimum $\chi^2$ point, which is taken to be the most likely configuration. The ellipse parameters were then used to determine the uncertainty on the Mylar and aluminum thicknesses.
From this procedure, we found the most likely source configuration to be \SI{40}{\nano\meter} of effective carrier material, \SI{6.50 \pm 2.04}{\micro\meter} of Mylar, and \SI{0.25(1.20:0.25)}{\micro\meter} of aluminum for $^{109}$Cd, and \SI{5}{\nano\meter} of effective carrier material, \SI{10.50(4.03)}{\micro\meter} of Mylar, and \SI{1.25(2.85:1.25)}{\micro\meter} of aluminum for $^{113}$Sn.

In order to correct for the energy loss due to the source configuration, we perform the peak fit for the 68\% confidence interval distribution corresponding to each source, for each pixel. 
The corrected central value is determined from the fit result for the minimum $\chi^2$ point. 
The uncertainty is taken to be maximum deviation between the central value and the points within the 68\% confidence interval. The results provide corrected \SI{}{\kilo e\volt} energy values for the lines used in the calibration scheme, for each source and 5 radial pixels.

For each pixel, we used the corrected energy values for both sources and performed a single linear calibration fit that accounted for both the statistical uncertainty (from the measured ADC energy values) and uncertainty due to the source configuration (from the corrected \SI{}{\kilo e\volt} energy values).
For the linear calibration gain uncertainty, we found that the contribution due to the uncertainty in the source configuration (on average \SI{5e-3}{ADC/\kilo e\volt}) dominates over that due to the statistical uncertainty (on average \SI{7e-4}{ADC/\kilo e\volt}).
But, for the linear calibration offset uncertainty, the contributions from the source configuration uncertainty (on average \SI{0.15}{ADC}) and the statistical uncertainty (on average \SI{0.12}{ADC}) were comparable.

\subsubsection{Calibration Results}
\label{subsec:calresults}

\begin{figure}
    \includegraphics{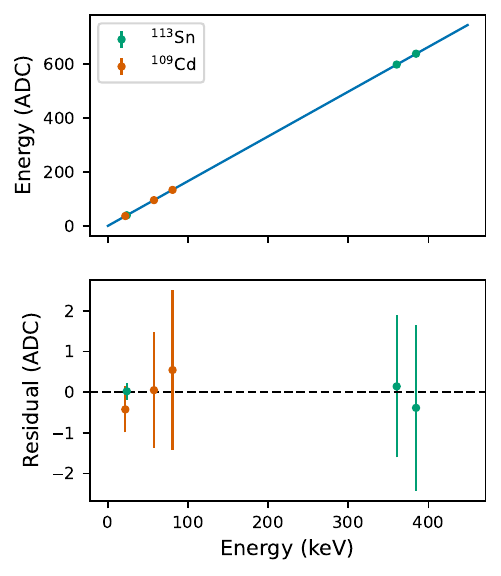}
    \caption{The top plot shows the combined $^{113}$Sn and $^{109}$Cd linear calibration result for pixel 76 from an x-ray and two conversion electron lines for each source (six total energy lines used in single calibration). The peak centroids in \SI{}{ADC} are obtained from a fit to the data. Centroid values in \SI{}{\kilo e\volt} are obtained from the fit results for the 68\% confidence interval distribution, from the convolved Geant4 simulated source configuration spectra. The bottom plot shows a residual plot. The $^{113}$Sn energy lines are shown in green, and the $^{109}$Cd lines are shown in orange.}
    \label{fig:Calibration_results_pix76}
\end{figure}

Following the calibration scheme presented in this section, we applied a temperature correction from \cite{PEHL196845} and an energy-loss correction for each pixel, constructed from the most likely source configuration.
After applying all corrections, the linear calibration gain parameter changed by $\sim0.9\%$ for all pixels, and the gain uncertainty decreased by a factor of $\sim1.9$ on average.
The linear calibration offset parameter increased on average by \SI{0.7}{ADC}, and the offset uncertainty decreased by a factor of $\sim5.4$ on average.
These results indicate that the applied corrections significantly reduce the calibration parameter uncertainties. The calibration results are shown in Fig.~\ref{fig:Calibration_results_pix76} for pixel 76 at $V_b=\SI{-300}{\volt}$. 
The top plot shows the calibration fit and the bottom plot shows the residuals to the linear model for each of the energy points used in the calibration. 

\begin{figure}
    \centering
    \includegraphics[width=\columnwidth]{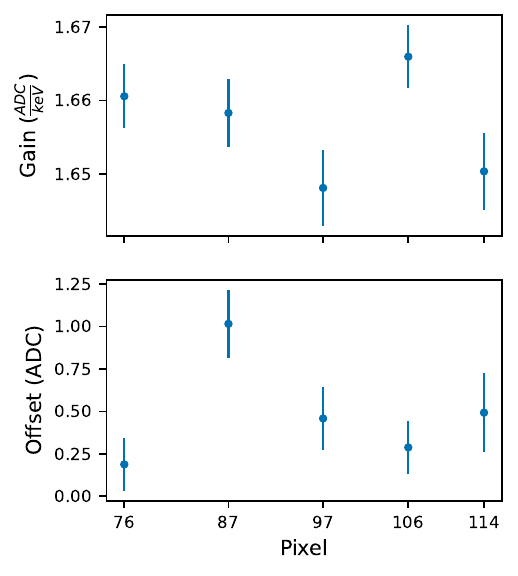}
    \caption{Linear calibration results for a radial row of pixels. The gain and offset parameters are shown in the top and bottom plots respectively.}
    \label{fig:Calibration_parameter_results}
\end{figure}

The calibration model parameters are shown in Fig.~\ref{fig:Calibration_parameter_results} for a radial row of pixels. 
Some variation in the calibration is expected since the pixels are connected to different combinations of electronic components, each of which may vary slightly.  
We note for Nab, the requirements for our calibration vary depending on the analysis method used to determine the beta-neutrino correlation and apply to the maximum uncertainties permitted for the extracted energy.  To date, the collaboration has focused on the analysis method presented in  Ref.~\cite{Fry_2019} with uncertainties listed in Table 1 (which determines the time of flight using parameterized fit functions for the magnetic field profiles).  
In this method, the calibration's linear gain is a fit parameter and it's uncertainty does not enter into the analysis. 
To meet the ultimate precision requirements of Nab, the offset parameter uncertainty should be $|\delta E_0| < \SI{0.3}{\kilo e\volt}$ ($\sim0.5$ ADC) for electron energy reconstruction, and as shown, all pixels analyzed meet this requirement.

\subsubsection{Non-linearity from Calibration Results}
\label{sub:non_linearity}

We expect the Si detector system response to be highly linear \cite{PEHL196845}, with our electronics chain evaluated to demonstrate smaller than 0.3\% non-linear response in  \cite{Broussard2017_2}, however non-linear effects can arise from charge-trapping in semi-conductor detectors \cite{Knoll_2024}.   
To achieve the target precision for the $a$ coefficient, the preferred analysis for Nab (see Ref.~\cite{Fry_2019}) requires the uncertainty in the maximum residual between a quadratic and linear calibrations to be $<\SI{1.5}{\kilo e\volt}$.
The uncertainties in the maximum residuals for our calibration were found to meet this specification over the full energy range of neutron beta decay except for pixel 114. Pixel 114 was located near the outer ring and had a maximum residual between the linear and quadratic calibrations near 195 keV with an uncertainty in the maximum residual of 2 keV. 
Given that the calibration source was placed near the center of the detector, this larger uncertainty is in part due to the larger distance and incident angle for pixel 114 relative to the source. 
The implication of this is that for ``production'' running in the Nab experiment, higher statistics calibration data sets will be required.
That being said, none of the pixels had statistically significant non-linear contributions and the uncertainty in possible non-linear effects did not significantly impact the work presented here.

\section{Proton Detection Characteristics}
\label{sec:proton_charac}

To characterize proton detection, the Manitoba facility was used to deliver \SI{25}{\kilo e\volt}, \SI{30}{\kilo e\volt}, and \SI{35}{\kilo e\volt} protons to the silicon detector system at a range of temperatures and reverse bias voltages $V_b$ (Sec.~\ref{sec:data_summary}).
The detection efficiency of the proton is strongly dependent on the magnitude of the ionization signal and the noise.
We present here the observed proton signal peak resolution with these detectors, and the stability of said peak with both temperature and time.

\subsection{Proton Energy Deposition}
\label{subsec:proton_peak_characteristics}

\begin{figure}
    \centering
    \includegraphics{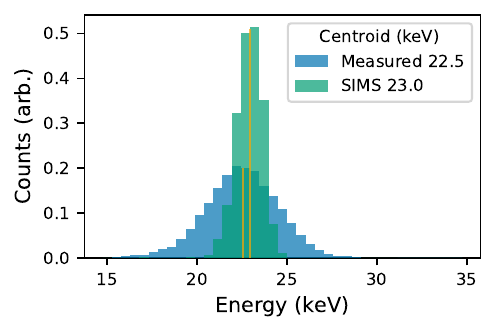}
    \caption{Normalized histograms of measured proton events as compared to a secondary-ion mass-spectroscopy (SIMS) based dead layer model of \SI{30}{\kilo e\volt} protons. Note that the protons lose approximately 6-\SI{7}{\kilo e\volt} through the entrance window of the detector. The SIMS model is simulated without noise.}
    \label{fig:SIMS}
\end{figure}

The trapezoidal filter parameters and calibration conversion coefficients were individually found for a set of pixels, so that the detected proton peak could be investigated precisely.
The \SI{25}{\kilo e\volt}, \SI{30}{\kilo e\volt}, and \SI{35}{\kilo e\volt} protons lost \SI{5.9\pm 0.25}{\kilo e\volt}, \SI{6.5\pm 0.25}{\kilo e\volt}, and \SI{6.8\pm 0.25}{\kilo e\volt}, respectively through the dead layer, and had full width at half maxima (FWHM) between 5 and \SI{7}{\kilo e\volt} (see Fig.~\ref{fig:SIMS}).
This is consistent with the FWHM and uncertainties found in Sec.~\ref{subsec:Source_Calibration}.
These losses were slightly less than the \SI{8.61 \pm 0.45}{\kilo e\volt} measured previously with similar 0.5 and \SI{1}{\milli\meter} thick detectors \cite{Salas-Bacci2014}.

Detected protons lose energy in the detector dead layer: the $p^+$ implantation region in the front face of the detector where charge collection efficiency is limited due to charge recombination (see Sec.~\ref{subsubsec:silicon_diode_detectors}).
The dead layer can be modeled as a region where no created charge is collected (a ``hard" dead layer) or one where, due to diffusive processes, partial charge collection occurs (a ``soft" dead layer)\cite{Wall2014,Hayen2023}.
The ionizing losses of this detector are consistent with a \SI{55\pm2}{\nano \meter} hard or \SI{60\pm2}{\nano \meter} soft dead layer (parametrized in Ref.~\cite{gugiattiCharacterisationSiliconDrift2020}), as determined from a simultaneous fit to the three proton energies.
This is consistent with the 50 to \SI{110}{\nano\meter} hard dead layer measured by Salas-Bacci et al. \cite{Salas-Bacci2014}.

A parameter free dead layer model was previously developed using input from manufacturer-provided secondary ion mass spectroscopy (SIMS) measurements of boron implantation into the detector face \cite{Hayen2023}.
This model also produced energy losses consistent with those observed, as shown in Fig.~\ref{fig:SIMS}.
If the SIMS data provided by the manufacturer is assumed to be accurate, the consistency of the SIMS dead layer model with measured proton energies suggests there are not energy losses from additional sources such as surface layers of water or hydrocarbons formed on the detector face as was observed in Ref.~\cite{Broussard2017_2}.

All three dead layer models do not account for channeling effects; however, for protons incident in a narrow angle range (corresponding to a subset of measured pixels), channeling can reduce the energy losses through the dead layer for a subset of pixels \cite{Dearnaley_1964,CZERBNIAK_1990, hoblerRandomChannelingStopping2006}.
A full characterization of the ionization energy loss mechanisms, including channeling, will require further study.

\subsection{Proton Peak Stability with Temperature}
\label{subsec:proton_peak_temp}

\begin{figure}
    \centering
    \includegraphics{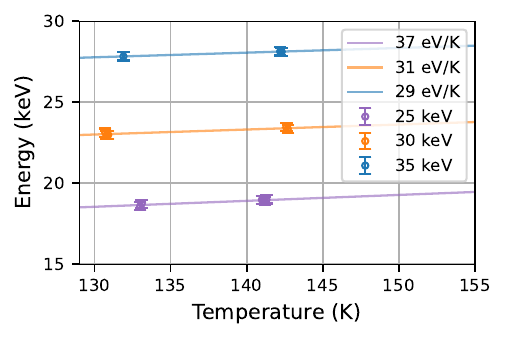}
    \caption{Measured proton peak centroids from pixel 76 as a function of temperature for three incident energies. Multiple datasets were taken at the respective temperatures and proton energies and the error bars come from the energy calibration ($\sim$~\SI{0.25}{\kilo e\volt}). Note that the legend contains both the slopes of the linear fit in \SI{}{e\volt\per\kelvin}, as well as the incident proton energies. Detected proton energy is along the vertical axis.}
    \label{fig:peak_versus_temp}
\end{figure}

As described in Section~\ref{subsubsec:temp_correct}, it is important to understand the movement of the proton peak with the temperature of the silicon wafer.
To this end, we logged the temperature of a sensor embedded in the detector's copper support enclosure.  The reported temperatures from the temperature logging program were interpolated to find the temperature corresponding to the timestamp of each event from the fast data acquisition within a proton dataset.
This was done because the temperature sensors were only read-out at \SI{1}{\second} intervals, and the proton events are uncorrelated in time.
Once the temperature was interpolated, the data was binned in \SI{0.1}{\kelvin} bins, and fit using a Gaussian and a linear background.
The centroids and errors from a representative pixel (pixel 76) are presented in Fig.~\ref{fig:peak_versus_temp}.
A linear fit to the \SI{30}{\kilo \eV} incident proton peak centroids shows a \SI{0.14\pm 0.18}{\% \per \kelvin} shift with the uncertainty in the slope arising from the calibration procedure.

\subsection{Proton Peak Stability Over Time}

Given the extended period over which Nab data will be taken, the stability of the system is a point of concern.
Effects like cryo-pumping can cause foreign material to condense on the surface of the silicon, creating an inconsistent deposit on top of the junction layers which further degrades the proton energy \cite{Broussard2017_2}.

\begin{figure}
    \centering
    \includegraphics{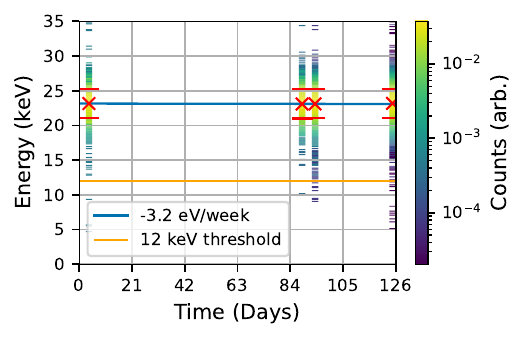}
    \caption{Two dimensional histogram of recorded proton energy versus time for \SI{30}{\kilo e\volt} incident energy protons, and detector temperatures between \SI{120}{\kelvin} and \SI{130}{\kelvin}. Here the counts have been normalized, and each dataset was fit with a Gaussian function to extract the centroid (red x's) and full width at half maximum (red error bars). The DAQ detection threshold for this pixel is \SI{12}{\kilo e\volt}. A linear fit is applied to the fit centroids to determine peak stability over time (in blue).}
    \label{fig:peak_versus_time}
\end{figure}

The calibration campaign discussed in this paper consists of data collected over a period of one year, allowing us to assess the stability for proton detection.
Figure~\ref{fig:peak_versus_time} shows an example of \SI{30}{\kilo e\volt} incident protons on one pixel in data collected over a 127 day period. 
The average stability for \SI{30}{\kilo e\volt} protons incident on a radial set of pixels was found to be well within our calibration uncertainties over this time.
Because each of the datasets collected were over a maximum period of twenty minutes, the temperature fluctuations within a given dataset are negligible compared to the calibration uncertainty.
Further, taking the number of standard deviations above threshold as a figure-of-merit for peak detection, we found the lower bound of the FWHM to be 2.5 $\sigma$ above the \SI{12}{\kilo e\volt} threshold for \SI{25}{\kilo e\volt} protons, 3.5-4.5 $\sigma$ for \SI{30}{\kilo e\volt} and 5.5-6.5 $\sigma$ for \SI{35}{\kilo e\volt}.

While installed in the Nab spectrometer, the detectors will be continuously cooled with a temperature-controlled gaseous helium system \cite{richburg2026}.
This differs from the procedure used in this publication, as the data collection process utilized liquid nitrogen.
A mass-flow controller was used to regulate a LN$_2$ supply to the detector, which remained cold for approximately one week at a time (until the LN$_2$ was spent).
This permitted fluctuations of $\sim$\SI{2}{\kelvin} over a three-day dataset, with additional systematic studies being collected during the warming period at the end.

The repeated cool-down and warm-up cycles offer an extreme case of cryo-pumping, during which contaminants from the vacuum chamber can condense on to the surface of the silicon.
We did not observe significant peak movement over this time (see Fig.~\ref{fig:peak_versus_time}).
The expected improvement to temperature stability in the Nab apparatus will further limit these effects.
Our primary goal is to demonstrate that these detectors provide reliable proton ionization signals under the operating conditions expected for Nab. 
The energy losses observed over the course of the measurements reported here were stable and consistent with the performance expected for Nab.

\section{Pulse Shape Analysis for Timing Bias Assessment for Nab}
\label{sec: Waveform Analysis}

The Nab experiment requires that the timing extraction is well understood to extract proton and electron timings and precisely determine the proton time of flight. 
To assess the timing performance of the detector we first analyzed the calibration source pulses to characterize the detector pulse shape response. 
We did this primarily through a comparison of the charge collection times between simulated and measured source events (parametrized by the pulse rise time)  to determine the detector electron drift velocity and Si bulk impurity density profile. 
Once we determined the pulse shape response we then used simulated pulses to investigate the timing capabilities of the detector.

\subsection{Pulse Shape Characterization}
\label{sec: Pulse shape characterization}

\subsubsection{Pulse Shape Simulation}
\label{sec:pulse_shape_simulation}

The pulse shape is determined by a convolution of the detector system electronics response and the induced current on the detector pixel contacts. 
We modeled the electronics response with a SPICE model, as developed in Ref.~\cite{Hayen2023}. 
The induced current is determined by the pixel weighting field and the electron and hole drift trajectories according to the Shockley-Ramo theorem \cite{Knoll_2024}. 
The former is the same for all inner pixels and is described in Ref.~\cite{Hayen2023}. 
Electron and hole drift trajectories are determined by the electric field within the detector, and the electron and hole drift mobilities.
The electric field is dependent on the bias voltage and impurity density:
\begin{equation}
\label{eq:field_strength}
|\bm{E}(z)| = \frac{V_b}{L}+\frac{NqL}{2\varepsilon}-\frac{Nq}{\varepsilon}z,
\end{equation}
where $N$ is the impurity density, $V_b$ is the bias voltage, and $\epsilon$ is the dielectric strength of silicon \cite{Knoll_2024}. 
In high-purity silicon, as used in Nab detectors, electron and hole mobilities are dominated by lattice scattering and are therefore not significantly affected by impurity density variations, but they have a power law temperature dependence \cite{Klaassen1992}.
The pulse shape is also dependent on the location within the detector where electron-hole pairs are created which is primarily dependent on incident particle energy and type.

The dominant pulse shape effects have been collected into a dedicated single detector simulation tool, NESSE (Nab Event Shape Simulation Effort) \cite{RjtaylNessePython}. 
The dead layer, aluminized grid, charge self repulsion, and charge recapture are estimated to have a negligible contribution to electron event rise times at the precision level of this study \cite{Hayen2023}.
The geometry of the detector contacts and impurity depositions, charge diffusion, energy deposition, electron-hole pair creation and scattering are fixed parameters that are described in detail in Ref.~\cite{Hayen2023}.
NESSE simulations handle these effects in the same way as described in Ref.~\cite{Hayen2023}, with incident particle energy depositions generated through the Geant4 simulation previously described.
The only varied simulation variables for pulse shape characterization were temperature, bulk impurity density, detector bias voltage, and mobility. 

The detector temperature and bias voltage were measured during data-taking. 
These values were used as input into the NESSE simulations.
Pulses were generated for a range of detector bias voltages ($V_b$): -300, -210, -180, -150, and \SI{-120}{\volt}.

The electron and hole drift velocities as a function of electric field strength were determined by an empirical fit to data from Ref.~\cite{Canali1975}, which was found to be most appropriate for simulations of Nab detectors due to the similar range of temperatures and electric field strengths used in the study and the Nab experiment \cite{Hayen2023}.
This data has an uncertainty of $\sim10\%$, so we tuned the drift velocities used in the NESSE simulations to our particular detector. 

The impurity density is expected to vary radially and axially in the silicon used in the Nab detectors because the silicon is produced using the ``float zone" method \cite{schroderSiliconGrownFloating2011a}.
An axial impurity density variation contributes to different average impurity densities between detectors. 
Radial variation in the impurity density causes variations between pixels in the same detector. 
The average bulk impurity density can be determined by the measured depletion voltage of the detector, but the radial gradients require a more localized characterization.
For the following studies pulses were generated with the impurity densities varied from \SI{0.2e10}{\per\cm\cubed} to \SI{4.2e10}{\per\cm\cubed} in steps of \SI{0.2e10}{\per\cm\cubed}.

The pulses were generated using energy deposition input from Geant4 simulations of monoenergetic \SI{30}{\kilo e\volt} protons, \SI{87}{\kilo e\volt} electrons from a $^{109}$Cd source, and \SI{364}{\kilo e\volt} electrons from a $^{113}$Sn source.
These are all events observed with the detector at the University of Manitoba proton source facility.
The stopping power of silicon is greater for protons than electrons, so electrons will penetrate deeper into the detector for the same given energy. 
Additionally for Nab the protons are accelerated to \SI{30}{\kilo e\volt}, but the electrons can have energies up to about \SI{1}{\mega e\volt} allowing them to penetrate as far as \SI{1.5}{\milli\meter} into the detector. 
The mobility in silicon of electrons is about five times greater than holes. 
Due to this mobility difference there is an approximately \SI{0.4}{\milli\meter} region extending from the detector entrance window in which the induced charge is electron transport dominated, meaning it takes the majority of electrons longer to reach the back contact than it takes holes to reach the front. 
Within this electron dominated region higher energy events typically have shorter rise times, since the electrons have a shorter distance to travel. 
Past this point higher energy events start to have longer rise times as the holes are generated farther away from the front face. 
All the events simulated for this study were electron-transport dominated.

\begin{figure}
    \centering
    \includegraphics{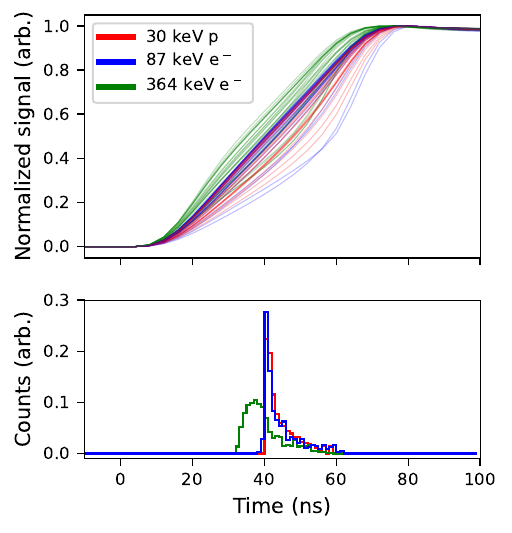}
    \caption{Simulated pulses from a \SI{30}{\kilo e\volt} proton beam with uniform pixel illumination, $^{109}$Cd \SI{87}{\kilo e\volt} conversion electrons, and $^{113}$Sn \SI{364}{\kilo e\volt} conversion electrons. 25 random pulses of each particle type are selected to show a representative sample of pulse shapes. All pulses are normalized to the pulse maximum to more easily compare pulse shapes. For all pulses, the impurity density is \SI{2e9}{\per\centi\meter\cubed}, the temperature is \SI{124.3}{\kelvin}, and the bias voltage is -\SI{150}{\volt}. Events located closer to the pixel edge have pulses with larger positive second derivatives and are less common. Below are presented normalized histograms of the times at which the pulses cross their 50\% threshold, shown in their corresponding colors. Color plot available online. }
    \label{fig:example-nesse-pulses}
\end{figure}

A random selection of 50 pulses normalized by pulse height are shown for \SI{30}{\kilo e\volt} protons, $^{109}$Cd \SI{87}{\kilo e\volt} conversion electrons, and $^{113}$Sn \SI{364}{\kilo e\volt} conversion electrons in Fig.~\ref{fig:example-nesse-pulses}. 
The samples were chosen from a pulse population uniformly distributed across the detector pixel.
Most pulses from the low energy events (30 and \SI{87}{\kilo e\volt}) have an approximately linear rising edge, but about one fifth of the pulses are incident near a pixel edge and have a significant curvature due to the weighting field in that area (see Ref.~\cite{Hayen2023} for more details). 
The higher energy Sn events are less linear due to a significant contribution to the pulse shape from hole transport, however the overall rise time is faster than the lower energy events.
From these events it is clear that even if temperature, voltage, and impurity density are constant, significant pulse shape differences are present for different energy events and different locations on the pixel simply due to the weighting potential.

In general, a shorter rise time (faster drift velocity) will decrease the timing offset, so one would want a high bias voltage, low impurity density, and low temperature to minimize the timing bias. 
The lowest temperature achieved by the cryogenic cooling system at the University of Manitoba was about \SI{120}{\kelvin}, and cooling performance is expected to be similar for the Nab experiment (see Sec. \ref{subsubsec:detectormount}). 
A temperature change of \SI{1}{\kelvin} will result in an electron drift velocity change of about 0.4\% in silicon near \SI{120}{\kelvin}, which causes a roughly equivalent shift in the rise times \cite{Canali1975}. 

\begin{figure}
    \centering
    \includegraphics{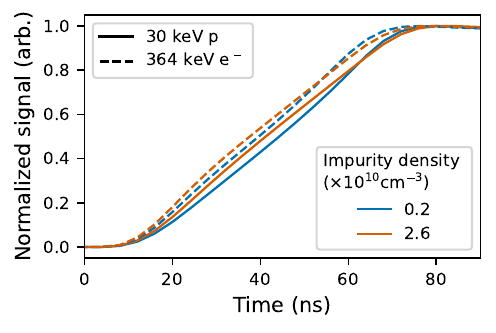}
    \caption{Averages of simulated pulses for a single uniformly illuminated pixel for different bulk impurity densities with a bias voltage of \SI{-150}{\volt}. Averages are shown for \SI{364}{\kilo e\volt} Sn conversion electrons (dashed) and \SI{30}{\kilo e\volt} protons (solid). }
    \label{fig:IDP-example}
\end{figure}

Figure~\ref{fig:IDP-example} shows the average simulated pulses over a single pixel for \SI{364}{\kilo e\volt} Sn conversion electrons and \SI{30}{\kilo e\volt} protons for two different impurity densities.
The average rise time difference due to this impurity density difference was 3.5\% and 5.4\% for the proton and Sn pulses respectively.
These shape differences will result in different timing offsets, $t_d$, for electron and proton events incident on pixels with different impurity densities. 

A higher bias voltage reduces the pulse rise time but also can increase the noise, which makes optimizing timing performance more difficult. 
One must choose a bias voltage that balances the two effects. 
A 0.2\% change in rise time per Volt is expected from the drift velocity dependence on electric field strength.

Changes of order 1 K in temperature, 10 V bias voltage, and $10^{10} \text{cm}^{-3}$  impurity density result in order 1\% changes in detector pulse rise times.
A precise account of the detector timing bias requires that all effects discussed above are understood and modeled together in the pulse shape simulations.

\subsubsection{Impurity Density Profile Characterization}

Large high purity silicon detectors can have significant radial variations in the impurity density. 
Due to the float zone process for manufacturing high purity silicon, impurities can concentrate in different radial regions of silicon wafer \cite{HAN2020125752, Hayen2023}.
The impurity density radial dependence of a detector can be non-monotonically increasing or decreasing, depending on the specifics of the manufacturing process and where along the silicon boule the wafer was taken from.
Therefore, for high precision experiments that utilize large high purity detectors, such as Nab, it may be necessary to measure the impurity density profile of each individual detector.
Here we describe a non-destructive method to determine the average impurity density of a pixel using calibration data and simulated pulses.

The detector mobilities and pixel impurity densities of the detector were determined by comparing 10-90\% rise time distributions between NESSE pulses and pulses from $^{109}$Cd and $^{113}$Sn sources (described in Sec.~\ref{subsec:Source_Calibration}).
The $^{113}$Sn \SI{364}{\kilo e\volt} and $^{109}$Cd \SI{87}{\kilo e\volt} conversion electron peaks were used in the rise time distribution comparison so that the calibration was performed over a relatively large energy range and at peaks with insignificant contributions from $\gamma$ rays. 
$\gamma$ rays penetrate the entire detector and therefore have different pulse shapes, which would complicate the analysis. 
The comparison was made for 5 pixels that extend radially out from the center (76, 87, 97, 106, and 114 as labeled in Fig.~\ref{fig:powered_pixels}). 
This was done to study the radial dependence of the detector impurity density profile. 

The NESSE rise time distributions were created by simulating 1000 random events for the $^{113}$Sn \SI{364}{\kilo e\volt} and $^{109}$Cd \SI{87}{\kilo e\volt} electrons on each of the five pixels at all voltages and impurity densities previously listed. 
The temperature of the simulation was matched to within \SI{1}{\kelvin} of the recorded detector temperatures. 

\begin{figure}
    \centering
    \includegraphics{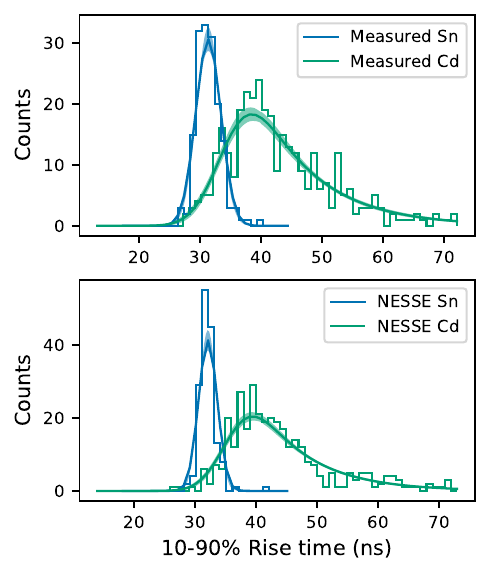}
    \caption{10-90\% rise time histograms from $^{113}$Sn \SI{364}{\kilo e\volt}  and $^{109}$Cd \SI{87}{\kilo e\volt} conversion electrons for measured data (top) and simulated with NESSE with additional noise taken from data (bottom) for a \SI{-300}{\volt} bias voltage. Lines are the maximum likelihood fit to a Gaussian and an exponentially modified Gaussian for the $^{113}$Sn and $^{109}$Cd data, respectively. Shaded regions are the $1\sigma$ error band.}
    \label{fig:risetime-distributions}
\end{figure}

Figure~\ref{fig:risetime-distributions} shows example rise time distributions from both Cd and Sn for both calibration and NESSE data.
The average rise time of a distribution was determined by fitting the 10-90\% rise time distributions of $^{113}$Sn \SI{364}{\kilo e\volt} to a Gaussian: 
\begin{equation}
    f(x)=\frac{1}{\sqrt{2\pi\sigma^2}}e^{-\frac{(x-\mu)^2}{2\sigma^2}}\text{,}
\end{equation} and of $^{109}$Cd \SI{87}{\kilo e\volt} to an exponentially modified Gaussian  \cite{grushkaCharacterizationExponentiallyModified1972}: 
\begin{equation}
    f(x)=\frac{1}{\tau\sigma\sqrt{2\pi}}\int_0^\infty e^{-\frac{(x-\mu-x')^2}{2\sigma^2}}e^{\frac{-x'}{\tau}}dx'\text{.}
\end{equation} 
The convolution of the exponential was included to match the right side tail in the $^{109}$Cd distributions.
The exponential towards increased rise times arises from the influence of noise on the rise time algorithm. 
Since the $^{109}$Cd electrons have about one quarter the energy of the $^{113}$Sn events, they likewise have one quarter the signal-to-noise ratio (SNR), and so the distribution broadens and the exponential tail appears. 
This was confirmed by adding representative noise sampled from the Manitoba data to the NESSE simulation pulses which recreates the exponential tail. 
The parameters of the functions above were used to estimate the distribution means. 
By the method of moments the mean of a Gaussian is $\mu$ and the mean of the exponentially modified gaussian is $\mu + \tau$. 
These values were extracted from the fits to determine the average 10-90\% rise time of the $^{113}$Sn and $^{109}$Cd data.

\begin{figure}
    \centering
    \includegraphics{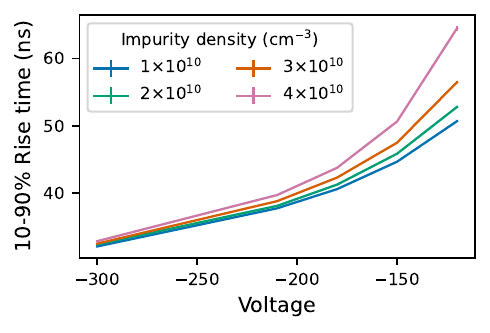}
    \caption{Average rise times from NESSE simulated pulses of $^{113}$Sn \SI{364}{\kilo e\volt} events on pixel 76 at \SI{122.3}{\kelvin}. Various impurity densities used for the n-type Si are drawn in different colors. Higher impurity densities cause larger average rise times at voltages closer to 0.}
    \label{fig:rt_voltage_nesse}
\end{figure}

The impurity density primarily affects signal risetimes through its impact on electric field strength (Eq.~\ref{eq:field_strength}). Higher impurity densities will have pulses with larger average rise times at low bias voltages. This effect can be seen clearly by plotting the average rise times as a function of the simulation bias voltage for various impurity densities. (Fig.~\ref{fig:rt_voltage_nesse}). Notably at large reverse bias voltages the average rise times for all impurity densities converge. These curves can be compared between simulation and data to determine the impurity density of each pixel, but first the electron mobility must be set, since a change in the simulation mobility can shift the entire curve up or down.

The rise times of the simulated pulses are affected by the drift velocity input into NESSE.
The charge collection time in the Nab silicon detectors from the \SI{87}{\kilo e\volt} electron from $^{109}$Cd and the \SI{364}{\kilo e\volt} electron from $^{113}$Sn are more affected by the drift electron mobility than by the hole mobility.
This allowed for a single scaling factor to be applied to the simulation rise times to account for the $\sim10\%$ uncertainty in the drift velocity data from Ref.~\cite{Canali1975}.
This scaling factor represents an adjustment to the electron drift velocity input into NESSE.
The scaling factor was determined by a simultaneous fit of both the $^{109}$Cd and $^{113}$Sn data to the full set of simulated average rise time data. 
The result was an overall scaling factor of $0.979\pm0.002$ relative to the rise times predicted using drift velocities from Ref.~\cite{Canali1975}.
Higher energy events would require a correction to both the electron and hole drift velocities.

\begin{figure}
    \centering
    \includegraphics{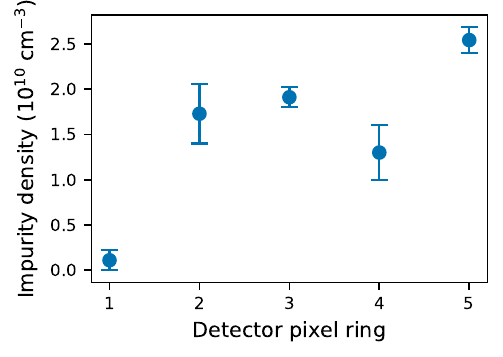}
    \caption{Fit results to impurity density of $^{113}$Sn \SI{364}{\kilo e\volt} and $^{109}$Cd \SI{87}{\kilo e\volt} events versus detector pixel ring, as described in the text. The error bars indicate one standard deviation.}
    \label{fig:IDP_fit}
\end{figure}

The impurity density concentration of each pixel was determined by a least squares fit of the simulated average rise times to both the average rise times of the $^{109}$Cd and $^{113}$Sn at all five voltages. 
The result of the fit is shown in Fig.~\ref{fig:IDP_fit}.
``Pixel ring" refers to which hexagonal ring the pixels tested are in; pixel 76 is in the first ring, 87 the second, etc. (see Fig.~\ref{fig:powered_pixels})
The shape of the impurity density gradient is roughly consistent with simulated impurity density gradients from Ref.~\cite{HAN2020125752}.
Ref.~\cite{HAN2020125752} describes how impurities can concentrate in different radial regions when silicon is produced using the float zone technique, causing multiple peaks in the impurity concentration profile, as seen in this detector. 
This range of impurity densities will have a significant impact on the shape of both proton and electron pulses (Fig.~\ref{fig:IDP-example}), and therefore has implications for the proton time of flight bias. 

\subsection{Timing Bias}
\label{sec:timing_bias}

As a proton time-of-flight measurement, the Nab experiment is particularly sensitive to a pulse timing bias. 
The experiment requires that the time the electron and proton hit the detector is known at the sub-ns level, but the detector and electronics response is at the \SI{10}{\nano \second} level.
An average shift in the extracted start times of pulses can introduce a bias in the measured electron-neutrino correlation in the Nab experiment \cite{Hayen2023}. 
We assign pulse times ($t_0$) using filters or fitting routines which are sensitive to the details of the pulse shape and the signal to noise ratio. 
If pulses had no noise then determining the ``true" start time ($t_T$) start time of the event would be straight-forward, but with a finite and low pulse to noise ratio, accurate timing is difficult.
Generally, any timing method will produce a time stamp that differs from the true incident time; this difference is called a timing offset ($t_d = t_0-t_T$).
Pulse shapes, and therefore the timing offsets, are different for protons and electrons.
This difference in timing offsets introduces a proton time of flight bias, $\Delta t_p$.
Therefore it is necessary to understand the differences in Nab pulse shapes to account for a possible time of flight bias. 

In this section we use the measured impurity density profile and drift velocity scaling factor from Section~\ref{sec: Pulse shape characterization} to simulate Nab pulses from the Manitoba detector. 
Since the start time, $t_T$, of simulated pulses is known, it is simple to determine $t_d$. 
We first describe our $t_0$ extraction algorithm, then the average timing bias for electrons and protons across the detector is determined.

\subsubsection{\texorpdfstring{$t_0$}{T0} extraction}
\label{subsec:t0-extraction}

Assigning a timestamp to a single event is difficult when the SNR is small ($\lesssim 5 {:} 1$), since the earliest trigger on the pulse must occur above the noise, and therefore after the pulse started rising. 
One approach to mitigate noise is to average many pulses together to determine an average pulse shape, but to do so without distorting the shape requires first precisely aligning the pulses with respect to time. 
To align the pulses properly we need good $t_0$ extraction first, therefore averaging cannot help us determine the pulse timing.
We applied a least squares fit of an exponentially modified sigmoid to pulses as a way to determine a pulse timestamp:
\begin{equation}
    f(x) = a + b x + c x^2 + d \frac{e^{-(x-t_0)/1250}}{1 + e^{-(x-t_0)/f}}.
\end{equation}
This extracted time ($t_0$) is the midpoint of the sigmoid rising edge, and is equivalent to the timing offset ($t_0=t_d$) since all simulated pulses have $t_T=0$.
This inherent timing offset is relatively large but the timing variance (or timing jitter) is similar to other timing extraction methods \cite{jezghaniRecursiveMethodRealTime2020}.

\subsubsection{Detector Timing Bias}

\begin{figure}
    \centering
    \includegraphics{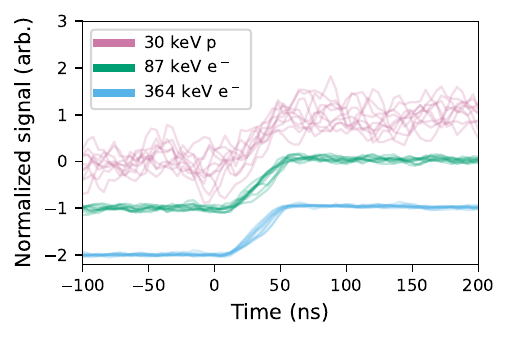}
    \caption{Ten randomly selected simulated pulses each from a \SI{30}{\kilo e\volt} proton beam with uniform pixel illumination, a $^{109}$Cd \SI{87}{\kilo e\volt} conversion electron, and a $^{113}$Sn \SI{364}{\kilo e\volt} conversion electron. Pulses simulated with a detector bias voltage of -300 V. All pulses have noise added that was sampled from pixel 76 at -300 V. Plotted pulses are then normalized to the pulse maximum.}
    \label{fig:example-noisy-pulses}
\end{figure}

The expected timing bias was determined by taking simulated electron and proton pulses, adding baseline noise from the corresponding calibration source data, and comparing the extracted $t_0$ to the simulation start time ($t_T=0$). 
Pulses were simulated for \SI{30}{\kilo e\volt} protons, \SI{87}{\kilo e\volt} electrons from a $^{109}$Cd source, and \SI{364}{\kilo e\volt} electrons from a $^{113}$Sn source as described in Section \ref{sec:pulse_shape_simulation}. 
The noise was sourced from pixel 76 baseline noise acquisition data at each bias voltage. 
The addition of noise was critical to the timing bias determination as it not only obscures the true start time but the overall pulse shape. 
Figure~\ref{fig:example-noisy-pulses} shows an example of each type of event with noise added. 
Compared to Fig.~\ref{fig:example-nesse-pulses} the event types and timings are much more difficult to accurately determine. 
The most prominent difference between the different types of events is that the SNR is significantly better for the higher energy events since they have larger pulse amplitudes, which makes $t_0$ extraction more precise. 

\begin{figure}
    \includegraphics{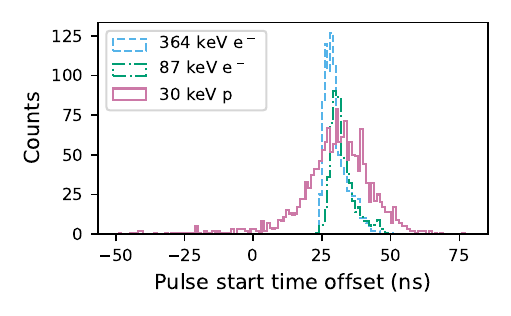}
    \caption{Measured pulse start time offsets of NESSE simulated pulses from a \SI{30}{\kilo e\volt} proton beam with uniform pixel illumination, a $^{109}$Cd \SI{87}{\kilo e\volt} conversion electron, and a $^{113}$Sn \SI{364}{\kilo e\volt} conversion electron, all with a -300 V detector bias voltage. Start time offset is determined by taking the difference from the true simulated incident time and the time extracted from an exponential logistic function fit to a pulse.}
    \label{fig:example-timing-histograms}
\end{figure}

Event timing offset distributions were extracted from NESSE pulses for impurity densities of \SI{0.2e9}, \SI{1.2e9},\SI{1.8e9},\SI{2e9}, and \SI{2.6e9}{\per\centi\meter\cubed} at -150, -210, and -300 V. An example for an impurity density of \SI{2e9}{\per\centi\meter\cubed} and a bias voltage of \SI{-300}{\volt} is shown in Fig.~\ref{fig:example-timing-histograms}
All event types had an average timing offset greater than the true start time, as expected from our $t_0$ extraction method. 
However, the different offset for each event type will result in an overall time of flight bias in the Nab experiment measurement.

\begin{figure}
    \includegraphics{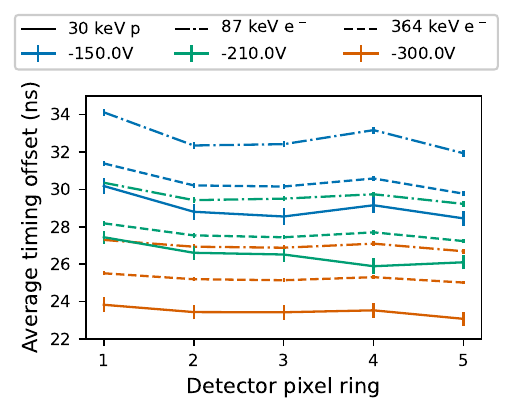}
    \caption{The average timing offset versus detector pixel ring is shown for NESSE simulated pulses from a \SI{30}{\kilo e\volt} proton beam with uniform pixel illumination (solid), a $^{109}$Cd \SI{87}{\kilo e\volt} conversion electron (dot-dashed), and a $^{113}$Sn \SI{364}{\kilo e\volt} conversion electron (dashed). Pixel ring represents which hexagonal ring the test pixels are in. The results are shown for three different voltages: -150 (blue), -210 (orange), and \SI{-300}{\volt} (green). Error bars correspond to the standard deviation of the timing offsets.}
    \label{fig:timing-bias}
\end{figure}

 The time of flight bias of a particular event is a result of the difference in pulse timing between the beta decay proton and electron. 
 We show the average timing offset for the three types of events versus pixel ring in Fig.~\ref{fig:timing-bias}, where the pixel ring corresponds to the pixels with impurity densities characterized in Section~\ref{sec: Pulse shape characterization}. 
 For all events, the timing offset average and variance was inversely proportional to the reverse bias voltage. 
 The shape of the impurity density gradient in Fig.~\ref{fig:IDP_fit} is mirrored in the shape of the pulse timings.
 The timing standard deviation across the detector for protons at \SI{-150}{\volt} was \SI{1.7}{\nano\second} and at \SI{-300}{\volt} was \SI{0.8}{\nano\second}.
 The largest difference in timing offsets evidently comes from the difference in incident particle energies; for example at a \SI{-300}{\volt} bias and the same impurity density the \SI{364}{\kilo e\volt} and \SI{87}{\kilo e\volt} electron events had an average timing offset difference of about \SI{2}{\nano\second}. 
 Such a difference in timing offsets will introduce an electron energy dependent proton time of flight bias. 
 If unaccounted for in the time of flight analysis these timing variations are above the allowable uncertainty ($< 0.3$ ns) for the Nab experiment.

 The uncertainty of the timing offsets at \SI{-300}{\volt} was on average \SI{0.2}{\nano\second}. 
 This suggests that the uncertainty in timing due to pulse shape differences at different locations on a pixel is below the Nab requirement, assuming all other timing effects have been accounted for. 

The above estimates for the timing bias uncertainty due to impurity density variations across the detector and incident locations in a pixel are expected to be larger than those for the Nab data.
The timing bias uncertainty is expected to improve due to decreased electronics noise from the updated  detector system electronics (see Section~\ref{subsubsec:Amplifier}).
It will also improve if better timing extraction methods, such as template fitting, are implemented. 
Template fitting is another common method to measure pulse features; it involves using Monte Carlo simulation to establish a template library from which pulses can be fit to, therefore determining energy and start times \cite{Knoll_2024}. 
We did not use template matching here since the data are from Monte Carlo simulations, but development of a full template library with NESSE may reduce Nab timing uncertainties significantly in the future.

If both the impurity density of each pixel and the timing response for the full range of electron energies can be determined, then the remaining Nab time-of-flight uncertainty attributable to the Si detector itself will be timing offset uncertainties like those determined here. 
For a detector biased at -300 V the proton offset uncertainty was on average 0.2 ns, limited by statistics.
The impurity density determination should improve as the detectors used in the Nab spectrometer will collect significantly more calibration data than what was used in this study.
Therefore the contribution to the time of flight uncertainty from the detector itself is expected to be less than \SI{0.2}{\nano\second}, below the overall \SI{0.3}{\nano\second} uncertainty requirement for the Nab precision goal that was discussed in Sec.~\ref{sec:nab_requirements}.
 
\section{Impact for the Nab Experiment}
\label{sec:operating_conditions}

The proton ionization signals were consistent with models for an approximately \SI{55}{\nano \meter} hard or \SI{60}{\nano \meter} soft dead layer and also a parameter-free dead layer model based on detector manufacturer-provided SIMS data.  
The detected proton energy was stable and no detector surface deposits were observed over roughly a year of repeated cryogenic cycling at pressures somewhat higher than those expected for the ultrahigh vacuum conditions of the Nab experiment.
Our preliminary assessment of the cross-talk for nearest-neighbor pixels suggests that it is below the percent-level, and should not present significant complications in the interpretation of proton signals.

A method for extracting the timing bias was presented which determines the inherent timing offset associated with signals from electron and proton events, for a given timing algorithm. 
By tying the analysis to a detailed model of the expected pulse shape and experimental noise, we provided accurate statistical predictions for the proton time-of-flight bias at sub-nanosecond levels.  
These estimates can be applied to simulations of the full Nab experiment, generalizing the results of the range of impurity densities and event topologies expected for Nab. 

The ability to predict the timing bias also provides us with guidance for the optimal running conditions for the Si detectors in Nab.  
Preliminary measurement demonstrated that electronic noise is roughly constant over the range of bias voltages between -200 and \SI{-320}{\volt}, with no significant change in the parallel noise observed due to leakage currents (from the analysis of  trapezoidal filter peaking time vs energy peak width). 
This suggests running the detectors at bias voltages near the electronics chain operation limit ($\sim300$V). 
Optimal signal-to-noise for the electronics is expected for detector assemblies near \SI{120}{\kelvin} \cite{Knoll_2024}, with lower temperatures resulting in higher drift velocities and reduced rise-times.  
We therefore expect operation of detectors with bias voltages near -300 V and temperatures near 120 K will provide near optimal performance for Nab.

The measurements presented here define expectations for proton detection consistent with the achievement of the ultimate precision goal, 0.1\% for the beta-antineutrino angular correlation.

\begin{acknowledgments}
This work was supported by the US Department of Energy (Contracts DE-FG02-97ER41042 and DE-AC05-00OR22725), and in part by the US Department of Energy, Office of Science, Office of Workforce Development for Teachers and Scientists (WDTS) Science Undergraduate Laboratory Internship (SULI) program, the National Science Foundation (Contracts 2209590, 2514991 and 2412782), and the National Science Engineering Research Council of Canada (Contracts NSERC-SAPPJ-2019-00043, NSERC-SAPPJ-2022-00024).

This research was supported through research cyber-infrastructure resources and services provided by the Partnership for an Advanced Computing Environment (PACE) at the Georgia Institute of technology, Atlanta, Georgia, USA. 

This research was also supported by the National Energy Research Scientific Computing Center (NERSC), a DOE Office of Science User Facility located at Lawrence Berkeley National Laboratory, operated under Contract No. DE-AC02-05CH11231 using NERSC award NP-ERCAPm2720.
\end{acknowledgments}

\bibliography{refs}

\end{document}